\documentclass[prd,twocolumn,amsmath,amssymb,nofootinbib,floatfix,superscriptaddress]{revtex4}

\usepackage{enumerate}
\usepackage{graphicx,bm}
\usepackage{placeins} 
\usepackage[normalem]{ulem} 

\makeatletter
\def\graphicscale{\twocolumn@sw{0.3}{0.4}}
\def\graphicthreescale{\twocolumn@sw{0.3}{0.4}}

\begin{document}

\title{Gauge fixing and gauge correlations in noncompact Abelian gauge models}

\author{Claudio Bonati} 
\affiliation{Dipartimento di Fisica dell'Universit\`a di Pisa
        and INFN Largo Pontecorvo 3, I-56127 Pisa, Italy}

\author{Andrea Pelissetto}
\affiliation{Dipartimento di Fisica dell'Universit\`a di Roma Sapienza
        and INFN Sezione di Roma I, I-00185 Roma, Italy}

\author{Ettore Vicari} 
\affiliation{Dipartimento di Fisica dell'Universit\`a di Pisa,
        Largo Pontecorvo 3, I-56127 Pisa, Italy}

\date{\today}

\begin{abstract}
We investigate some general properties of linear gauge fixings and gauge-field
correlators in lattice models with noncompact U(1) gauge symmetry. In
particular, we show that, even in the presence of a gauge fixing,
 some gauge-field observables
(like the photon-mass operator) are not well-defined, depending on the
specific gauge fixing adopted and on its implementation. 
Numerical tests carried out in the three-dimensional noncompact 
lattice Abelian Higgs
model fully support the analytical results and provide further insights.
\end{abstract}

\maketitle

\section{Introduction}\label{sec:intro}

Some nonperturbative features of quantum field theories (QFTs) can be studied
from first principles by using the lattice discretization. In this formulation
the Euclidean version of the theory is regularized on a space-time lattice, and
the QFT problem is mapped to a statistical-mechanics one. Continuum physics
emerges as the correlation length of the statistical system diverges, i.e.,
close to a continuous phase transition (critical point) of the lattice
system. For this strategy to be feasible, there should exist a stable  fixed
point of the QFT renormalization group (RG) flow, which encodes the universal
properties of the critical point of the statistical system.

This approach has been extensively used to investigate for example
four-dimensional non-Abelian gauge theories and QCD in particular
\cite{MM_book, DGDT_book}, and the $\phi^4$ QFTs associated with classical and
quantum phase transitions in lower-dimensional systems \cite{ZJ_book,
Pelissetto:2000ek, Sachdev_book}. Only in a few cases 
is it possible to carry out
this strategy with full analytical control \cite{Seiler_book,GJ_book}, so that 
one has to rely on numerical simulations of the discretized theory. 

Four-dimensional non-Abelian gauge theories are peculiar, since the existence
of a fixed point of the RG flow can be shown analytically by using one-loop
perturbation theory \cite{Gross:1973id, Politzer:1973fx, Coleman:1973sx}. For
typical three-dimensional QFTs this is not the case, and the fixed point, if
present, is generically in the strongly coupled regime.  To analytically
investigate its existence and extract universal information, nonperturbative
approaches are required, like the expansion in the number of components
\cite{MZ-03} or the continuation in the number of dimensions obtained by
resumming the $\epsilon$-expansion series \cite{Wilson:1973jj, ZJ_book}.
 
Three-dimensional gauge theories coupled to matter fields have features in
common both with four-dimensional non-Abelian gauge theories and with
three-dimensional scalar models. On the one hand, the gauge coupling of
three-dimensional gauge theories has positive mass dimension (the theory is
super-renormalizable), thus the energy scaling of the coupling is dictated by
dimensional analysis, and asymptotic freedom is clear already at tree level. On
the other hand, there is also the possibility that nontrivial fixed points
exist, at which the gauge coupling does not vanish, and which are usually
referred to as charged fixed points. While the asymptotically free fixed points
of three-dimensional Abelian and non-Abelian gauge theories have been
thoroughly investigated by numerical simulations (see, e.g.,
Refs.~\cite{Bhanot:1980pc,Caselle:2014eka,Athenodorou:2018sab,%
Teper:1998te,Athenodorou:2016ebg,Bonati:2020orj}), the case of the charged
fixed points has attracted less attention until quite recently, when the
existence of strongly coupled charged fixed points has been suggested to
explain some peculiar critical phenomena \cite{SBSVF-04, Fradkin_book,
Sachdev:2018ddg, Moessner_book}. 

The existence of these charged fixed points, and
their critical properties, can be investigated using several complementary
techniques: the $\epsilon$ expansion close to four dimensions~\cite{HLM-74,
FH-96, IZMHS-19, Das:2018qmx, Sachdev:2018nbk, Bonati:2021tvg, Bonati:2021rzx,
fermioni1, fermioni2, fermioni3}, the expansion in the number of
components~\cite{ZJ_book, MZ-03, fermioni4} and numerical simulation of lattice
models.  Numerical studies have recently addressed this issue in the
Abelian-Higgs (AH) model, i.e., in scalar quantum electrodynamics with
$N$-component scalar fields, and there is by now compelling evidence that some
lattice models undergo a continuous transition related to the AH QFT charged
fixed point. This has been observed for $N\gtrsim 7$ using the noncompact
discretization \cite{Bonati:2020jlm, Bonati:2022ifi} and the higher-charge
compact discretization \cite{Bonati:2020ssr, Bonati:2022oez}, while only
first-order phase transitions have been found in other cases
\cite{Pelissetto:2019zvh, Pelissetto:2019iic, Pelissetto:2019thf,
Bracci-Testasecca:2022mxc} and for smaller $N$ values \cite{Bonati:2020jlm,
Bonati:2020ssr, MV-08, KMPST-08}.  For $N=2$, continuous transitions were
observed \cite{Pelissetto:2019zvh, Pelissetto:2019thf}, where gauge fields play
no role. Topological excitations likely play an important role for the existence
of the charge fixed point; however, this point is not yet fully understood
\cite{MV-04, MS-90, SP-15, Pelissetto:2020yas, Bonati:2022srq}.

Analytical and numerical results thus support the fact that 
gauge-invariant correlators are well defined beyond perturbation theory in the
Abelian Higgs QFT, if a large enough number of scalar flavors is present. It
is then natural to ask if gauge-dependent correlators, which play a fundamental
role in the usual perturbative treatment of gauge QFTs, can be given a similar
nonperturbative status.

In this work we aim to investigate this point and, more generally,
to clarify how the large-distance behavior of the gauge-field correlators
depends on the gauge-fixing procedure adopted. For this purpose, we study gauge
correlations in the noncompact formulation of Abelian gauge models. The
relations that will be derived are independent of the matter content of the 
theory.
Moreover, they are valid in the whole phase diagram of the model, 
and not only on the critical lines associated with charged fixed points. 
In the present paper, we also add a numerical study of 
the behavior of the gauge-field
correlations  in generic points of the phase diagram of the three-dimensional
Abelian-Higgs model, which provides further insights on the role of the 
different gauge fixings.
A detailed analysis of the critical behavior 
is left to a forthcoming  paper 
\cite{BPV-in-prep}.

To make gauge correlation functions well-defined, it is necessary to
introduce a gauge-fixing term, that completely breaks the gauge invariance of
the model. In noncompact discretizations, the gauge fixing plays a crucial
role, since, only in the presence of a gauge fixing, the partition function and
the average values of nongauge-invariant quantities are finite. This is at
variance with what happens in compact formulations, in which a gauge fixing is
not necessary. Also in the absence of it, the partition function is
well-defined and so are average values of nongauge invariant quantities. In
particular, correlations of nongauge invariant quantities are either trivial or
equivalent to gauge-invariant observables, obtained by averaging the nongauge
invariant quantity over the whole (compact) group of gauge transformations
\cite{Elitzur:1975im, DeAngelis:1977su, IZ_book}. The latter equivalence does
not hold in noncompact formulations, since the group of gauge transformations
is not compact and therefore, averages over all gauge transformations are not
defined.  

Once a gauge fixing is introduced, the first point to be investigated is
whether and how results for nongauge-invariant quantities depend on it.  Here
we consider two widely used gauge fixings, the axial and the Lorenz one. We
derive general results and perform a complementary numerical study in the AH
model. They both indicate that gauge correlations depend somehow on the gauge
choice made. In particular, we show that the photon-mass operator is
well-defined only in what we call the hard Lorenz gauge (see Sec.~\ref{sec:2}).
Unphysical results are obtained when using the axial gauge and the soft Lorenz
gauge. The conclusions of this work should be independent of the type of matter
fields considered (fermions or bosons) as they only rely on some specific
features of the gauge fixing functions.

The paper is organized as follows. In Sec.~\ref{sec:2} we introduce the lattice
model, define the gauge fixings and the gauge observable that we will focus on.
In Sec.~\ref{sec:3} we derive general relations, which are independent of the
nature of the matter fields, between the gauge field correlation functions in the
presence of different gauge fixings. In Sec.~\ref{sec:res} we present numerical
results obtained in the scalar AH model, with the purpose of determining the
behavior of gauge field correlation functions in the different phases present in the
model. In Sec.~\ref{sec:FT} we review some field-theory results for the gauge
dependence of the gauge field correlation functions. Finally, in Sec.~\ref{sec:concl}
we draw our conclusions. In App.~\ref{sec:A}, we summarize some analytic
results for the pure gauge model, while in App.~\ref{sec:B} we derive some
general relations for the gauge-dependent part of the gauge correlation
functions.

\section{The lattice model}\label{sec:2}

We consider a noncompact Abelian gauge theory on a $d$-dimensional cubic-like
lattice of size $L$, with fermionic and bosonic matter fields that we
collectively indicate with $\Psi$ and $\Phi$, respectively. The gauge
interaction is mediated by real fields $A_{{\bm x},\mu}\in\mathbb{R}$, defined
on the lattice links, each link being labeled by a lattice site ${\bm x}$ and
a positive lattice direction $\hat{\mu}$ ($\mu=1,\ldots,d$).  The action is
given by
\begin{equation}
S  = S_{\rm matter}(\Psi,\Phi,A) + S_{\rm gauge}(A), 
\end{equation}
where $S_{\rm matter}$ is the action for the matter fermionic and bosonic
fields, and $S_{\rm gauge}(A)$ is the action for the gauge fields, 
which is given by
\begin{equation}
S_{\rm gauge}(A) = 
\frac{\kappa}{2} \sum_{{\bm x},\mu>\nu} (\Delta_{\mu} A_{{\bm x},\nu} -
\Delta_{\nu} A_{{\bm x},\mu})^2\, .
\end{equation}
Here  $\kappa$ is the inverse lattice gauge coupling, $\Delta_\mu$ is a
discrete derivative defined by
$\Delta_\mu f_{{\bm x}} = f_{{\bm x}+\hat{\mu}}- f_{{\bm x}}$,
and we have taken the lattice spacing equal to one.  We assume that the action is
invariant under local gauge transformations, which act on the gauge field as 
\begin{equation}
A_{{\bm x},\mu} \to A_{{\bm x},\mu}' = A_{{\bm x},\mu}  - \Delta_\mu \phi_{\bm x} \; .
\label{eq:gaugetrans}
\end{equation}
Matter fields do not couple with $A_{{\bm x},\mu}$ directly, but rather through
$\lambda_{{\bm x},\mu} = \exp(i A_{{\bm x},\mu})$. This implies that, in a
finite system with periodic boundary conditions, the action $S$ is also
invariant under the transformation $A_{{\bm x},\mu}\to A_{{\bm x},\mu} + 2\pi
n_{\mu}$, where $n_{\mu}\in\mathbb{Z}$ depends on the direction $\mu$ but not
on the point ${\bm x}$. This transformation makes the averages of some
gauge-invariant quantities (for instance, of Polyakov loops, which, in
noncompact formulations, are defined as the sum of the gauge fields along paths
that wrap around the lattice), ill defined. To make the averages of all
gauge-invariant observables well-defined on a finite lattice, we adopt $C^*$
boundary conditions \cite{Kronfeld:1990qu, Lucini:2015hfa, Bonati:2020jlm},
that    correspond to considering antiperiodic boundary conditions for the
gauge fields, i.e., to 
\begin{equation}
A_{{\bm x} + L\hat{\nu},\mu}= -A_{{\bm x},\mu}\ ,
\end{equation}
for all lattice directions $\nu$.  When using $C^*$ boundary conditions, the
local U(1) gauge symmetry is preserved by using antiperiodic gauge
transformations $\phi_{\bm x}$ in Eq.~\eqref{eq:gaugetrans}.

To study correlation functions of the gauge fields, it is necessary to add a
gauge fixing. We consider gauge fixings that are linear in the fields and that
are translation invariant. We introduce a gauge-fixing function 
\begin{equation}
   F_{\bm x}(A) = \sum_{{\bm y}\mu} 
   M_{{\bm x} - {\bm y},\mu} A_{{\bm y},\mu},
\label{GF-M}
\end{equation}
where $M_{{\bm x},\mu}$ is a field-independent vector, and define the partition
function as
\begin{equation}
  Z_{\rm hard} = \int [\mathrm{d}\Phi\mathrm{d}\bar{\Phi}]
       [\mathrm{d}\Psi\mathrm{d}\bar{\Psi}][dA] \; \left[\prod_{\bm x} 
  \delta[F_{{\bm x}}(A)]\right]\, e^{-S},
\label{Zhard}
\end{equation}
where the product extends to all lattice sites.  Note that the insertion of the
gauge-fixing term does not change the expectation values of gauge-invariant
quantities.  In perturbation theory, one usually replaces the partition
function (\ref{Zhard}) with a different one (see, e.g., Refs.~\cite{MM_book,
ZJ_book, R_book}), defined by adding a term of the form 
\begin{equation}
      S_{\rm GF}(A) = {1\over 2\zeta} \sum_{\bm x}   [F_{{\bm x}}(A)]^2
\label{def-Gfunction}
\end{equation}
to the action. In this case one considers 
the partition function
\begin{equation}
  Z_{\rm soft} = \int [\mathrm{d}\Phi\mathrm{d}\bar{\Phi}]
       [\mathrm{d}\Psi\mathrm{d}\bar{\Psi}][dA] \;
    e^{-S - S_{\rm GF}(A)}.
\label{Zsoft}
\end{equation}
Since the gauge-fixing function is linear in the gauge fields, no
field-dependent Jacobian should be considered in the gauge-fixed model and,
therefore, no Faddeev-Popov term should be added. The partition function
$Z_{\rm soft}$ depends on the parameter $\zeta$.  For $\zeta\to 0$, the model
with partition function  (\ref{Zsoft}) is equivalent to the one with partition
function (\ref{Zhard}). We will call the gauge fixings appearing in
Eqs.~(\ref{Zhard}) and (\ref{Zsoft}) hard and soft gauge fixing, respectively.

In this work we will mainly focus on two widely used gauge fixing functions.
We consider the axial gauge fixing with
\begin{equation} 
F_{A,{\bm x}}(A) = A_{{\bm x},d}, 
\label{eq:axialgauge}
\end{equation}
and the Lorenz gauge fixing with 
\begin{equation} 
F_{L,{\bm x}}(A) = \sum_{\mu=1}^d (A_{{\bm x},\mu} - A_{{\bm x}-\hat{\mu},\mu}).
\label{eq:Lorenzgauge}
\end{equation}
Note that in a finite system with $C^*$ boundary conditions, both gauge fixings
completely fix the gauge (they are {\em complete} gauge fixings). Indeed, there
are no distinct configurations $A_{{\bm x},\mu}$ and $A_{{\bm x},\mu}'$ related
by a gauge transformation such that $F_{{\bm x}}(A) = F_{{\bm x}}(A') = 0$ for
all lattice points ${\bm x}$.

We consider correlation functions of the gauge fields. We define the Fourier
transform of the field as~\footnote{The added factor $e^{i p_\mu/2}$ is needed
to guarantee that $\widetilde{A}_{\mu}(\bm p)$ is odd under reflections in momentum
space, ${\bm p} \to (p_1,\ldots,-p_\mu,\ldots,p_d)$. Intuitively, it can be
understood by noting that $A_{{\bm x},\mu}$ is associated with a lattice link
and thus it would be more naturally considered as a function of the link
midpoint, i.e., we should write it as $A_{{\bm x}+\hat{\mu}/2,\mu}$. }
\begin{equation}
\widetilde{A}_{\mu}(\bm p)= 
e^{i p_\mu/2} \sum_x A_{{\bm x},\mu}e^{i{\bm p}\cdot{\bm x}}\, .
\end{equation}
Under $C^*$ boundary conditions, 
$A_{{\bm x},\mu}$ is antiperiodic, so that the allowed momenta
for $\widetilde{A}_{\mu}(\bm p)$
are ${\bm p} = (2 n_1 + 1, \ldots , 2 n_d + 1) \pi/L$ ($n_i = 0,\ldots L-1$).
In particular, ${\bm p} = 0$ is not an allowed momentum.
The corresponding momentum-space two-point function is
\begin{equation}
 \widetilde{G}_{\mu\nu}({\bm p})=
 \frac{1}{L^d}\langle \widetilde{A}_\mu(\bm p)\widetilde{A}_\nu(-\bm p)\rangle
   \,. \\
\end{equation}
We assume that the matter action is invariant under charge conjugation.
As this property is preserved by the $C^*$ boundary conditions
and by linear gauge fixings, the full theory is also invariant under charge
conjugation, which guarantees $\langle A_{{\bm x},\mu}\rangle=0$.

We also consider the composite operator 
\begin{equation}
    B_{\bm x} = \sum_\mu A^2_{{\bm x},\mu},
\end{equation}
which, in perturbative approaches, is included in the action to provide a mass
to the photon and therefore an infrared regulator to the theory (see, e.g.,
Ref.~\cite{ZJ_book}).  We define its Fourier transform
\begin{equation}
\widetilde{B}(\bm p)=\sum_x B_{\bm x} e^{i{\bm p}\cdot{\bm x}}\, ,
\end{equation}
where ${\bm p} = (2 n_1, \ldots , 2 n_d) \pi/L$ ($B_{\bm x}$ is periodic)
and the correlation function
\begin{equation}
 \widetilde{G}_B({\bm p})=\frac{1}{L^d}\left[\langle \widetilde{B}(\bm p)\widetilde{B}(-\bm p)\rangle
-\langle \widetilde{B}(\bm p)\rangle\langle\widetilde{B}(-\bm p)\rangle \right]\,.
\end{equation}
The long-distance properties of the correlators $\widetilde{G}_{\mu\nu}(\bm p)$ and
$\widetilde{G}_B(\bm p)$ can be determined by studying the gauge susceptibilities
\begin{equation}
\chi_{\mu\nu} = \widetilde{G}_{\mu\nu}({\bm p}_a)\,,\quad  
\chi_B = \widetilde{G}_B({\bm 0})\ , 
\label{def-chi}
\end{equation}
where the momentum ${\bm p}_a$ is defined by
\begin{equation}\label{eq:pmin}
{\bm p}_a=(p_{\rm min}, \ldots, p_{\rm min})  \qquad p_{\rm min} = {\pi\over L}
\; .
\end{equation}
Note that, since $A_{{\bm x},\mu}$ is antiperiodic, each component of the 
momentum can ony take the values $(2 n + 1) p_{\rm min}$ and thus ${\bm p}_a$ is 
one of the acceptable momenta for which $|{\bm p}|$ is as small as possible.

\section{Correlation functions in different gauges}\label{sec:3}

In this section we derive relations among correlation functions in different
gauges. These relations will help us to understand the nonperturbative behavior
of correlation functions, that will be discussed in Sec.~\ref{sec:res}.  We
focus on the axial and Lorenz gauge, but it is easy to generalize the
discussion to any arbitrary gauge-fixing function that is linear in the gauge
field. Moreover, all results concerning the gauge-field two-point correlation
functions can in principle be generalized to any correlation function of
the gauge fields. Finally, note that all results are 
independent of the nature of the matter fields.

\subsection{Hard Lorenz and axial gauges}

To relate Lorenz-gauge and axial-gauge results, we first determine 
a gauge transformation that maps the Lorenz gauge fixing onto the 
axial one. More precisely, given a field configuration $\{A_{{\bm x},\mu}\}$
we want to determine a gauge transformation (\ref{eq:gaugetrans}),
i.e. a function $\phi_{\bm x}$, such that
\begin{equation}
   {A'}_{{\bm x},d} = \sum_{\mu} (A_{{\bm x},\mu} - A_{{\bm x}-\hat{\mu},\mu}).
\end{equation}
Working in Fourier space, this corresponds to choosing 
\begin{equation}
    \widetilde{\phi}({\bm p}) = {i\over \hat{p}_d} 
  \left(i e^{i p_d/2} \sum_\mu \hat{p}_\mu \widetilde{A}_\mu({\bm p}) + 
      \widetilde{A}_d({\bm p}) \right),
\label{defphi}
\end{equation}
where $\hat{p}_\mu = 2 \sin (p_\mu/2)$.  This transformation is well defined on
a finite lattice with $C^*$ boundary conditions as $\hat{p}_d$ never vanishes.
It maps the action with a soft Lorenz gauge fixing onto the axial-gauge action
with the same parameter $\zeta$.  If we take the limit $\zeta\to0$, it allows
us to relate the two hard gauge-fixed models. 

To relate correlation functions we interpret the gauge transformation with
gauge function (\ref{defphi}) as a change of variables. Since the
transformation is linear in the fields, the Jacobian is independent of the fields and plays no role. Therefore,
if $O(A_{{\bm x},\mu})$ is a gauge-dependent operator, we have
\begin{equation}
\left\langle O(A_{{\bm x},\mu}) \right\rangle_{A,\zeta} = 
\left\langle O(A_{{\bm x},\mu} - \Delta_\mu\phi_{\bm x} ) 
\right\rangle_{L,\zeta}
\label{relations-A-L}
\end{equation}
where $\phi_{\bm x}$ is the anti-Fourier transform of Eq.~(\ref{defphi}) and the
two average values refer to the models with axial (A) and Lorenz (L) 
soft gauge fixing, respectively, with the same parameter $\zeta$. 

We can use Eq.~(\ref{relations-A-L}) to relate 
$\widetilde{G}^{(A)}_{\mu\nu}({\bm p})$ and $\widetilde{G}^{(L)}_{\mu\nu}({\bm p})$ 
(axial and Lorenz gauge, respectively). Considering only the hard case
($\zeta = 0$), 
using $\sum_\mu \hat{p}_\mu \widetilde{G}^{(L)}_{\mu\nu}({\bm p}) = 0$ 
(see App.~\ref{sec:B}),
we can express $\widetilde{G}^{(L)}_{d \mu}({\bm p})$ in terms of 
the components of the Lorenz function 
$\widetilde{G}^{(L)}_{\mu\nu}({\bm p})$ with $\mu,\nu\le (d-1)$.
This allows us to prove the relation ($1\le \mu,\nu\le d-1$) ,
\begin{equation}
\label{GA-GL} 
\begin{aligned}
\widetilde{G}^{(A)}_{\mu\nu}({\bm p}) =  \widetilde{G}^{(L)}_{\mu\nu}({\bm p}) + 
     {\hat{p}_\mu\hat{p}_\nu\over \hat{p}^4_d} 
     \sum_{\alpha\beta} \hat{p}_\alpha\hat{p}_\beta 
     \widetilde{G}^{(L)}_{\alpha\beta}({\bm p})+ \\
+ {\hat{p}_\mu\over \hat{p}^2_d}
     \sum_{\alpha} \hat{p}_\alpha \widetilde{G}^{(L)}_{\alpha\nu}({\bm p})   
     + {\hat{p}_\nu\over \hat{p}^2_d}
     \sum_{\alpha} \hat{p}_\alpha \widetilde{G}^{(L)}_{\alpha\mu}({\bm p}) \;,
\end{aligned}
\end{equation}
where $\alpha$ and $\beta$ run from 1 to $(d-1)$ only.  Obviously, as we are
considering the hard gauge fixing, $\widetilde{G}^{(A)}_{\mu\nu}({\bm p}) = 0$, if
$\mu$ or $\nu$ are equal to $d$.  We can use Eq.~(\ref{GA-GL}) to relate
$\chi_{\mu\nu}^{(A)}$ with $\chi_{\mu\nu}^{(L)}$. Because of the cubic
symmetry of the lattice and of the momentum ${\bm p}_{\rm a}$ (see Eq~\eqref{eq:pmin}), 
only two components of $\widetilde{G}^{(L)}_{\mu\nu}({\bm p}_a)$ are
independent. Therefore, we can write
\begin{equation}
   \chi^{(L)}_{\mu\nu} = 
   \widetilde{G}^{(L)}_{\mu\nu}({\bm p}_a) = 
    c_{L1} \delta_{\mu\nu} + c_{L2} (1 - \delta_{\mu\nu}),
\label{Lorenz-parametrization}
\end{equation}
where $ c_{L2} = - c_{L1}/(d-1)$ because of the Lorenz condition (see
App.~\ref{sec:B}).  Substituting in Eq.~(\ref{GA-GL}), we obtain (again $1\le
\mu,\nu\le d-1$):
\begin{equation}
   \chi^{(A)}_{\mu\nu} = 
 \widetilde{G}^{(A)}_{\mu\nu}({\bm p}_a) = c_{A1} \delta_{\mu\nu} + 
      c_{A2} (1 - \delta_{\mu\nu}),
\end{equation}
with
\begin{equation}
c_{A1} = {2 d \over d-1} c_{L1} , \qquad
c_{A2} = - d c_{L2}.
\label{chiHA-chiHL}
\end{equation}
The simple relations (\ref{chiHA-chiHL}) and 
(\ref{GA-GL}) do not extend, however, to composite operators.  
Indeed, the
transformation with function (\ref{defphi}) that relates the two gauges is
singular in the limit $L\to \infty$, because of the factor $1/\hat{p}_d$, which
diverges as $L\to\infty$.  This shows up in the presence of singular
coefficients in Eq.~(\ref{GA-GL}).  As a consequence, as we discuss in
Sec.~\ref{sec:res}, the average 
\begin{equation}
   \langle B_{\bm x} \rangle = {1\over V} \sum_{\mu\nu} 
   \sum_p \widetilde{G}_{\mu\nu}({\bm p})
\label{B-sum}
\end{equation}
behaves differently in the axial and Lorenz gauges.

\subsection{Hard and soft axial gauges}

Let us now determine how correlation functions vary in soft axial gauges as the
parameter $\zeta$ varies. As before, we consider changes of variables that are
gauge transformations. For the case at hand, we consider the gauge function
\begin{equation}
 \widetilde{\phi}({\bm p}) = \left( 1 - \sqrt{\zeta_2\over \zeta_1}\right) 
   {i\over \hat{p}_d} \widetilde{A}_d({\bm p}),
\end{equation}
that allows us to map the model with parameter $\zeta_1$ onto the model with
parameter $\zeta_2$. It is immediate to relate correlation functions. Using
Eq.~(\ref{relations-A-L}) modified for the case at hand, we obtain 
($\mu,\nu\le d-1$)
\begin{equation}
\begin{aligned}
& \widetilde{G}^{(A)}_{\mu\nu}({\bm p},\zeta_2) = 
    \widetilde{G}^{(A)}_{\mu\nu}({\bm p},\zeta_1) + 
    r^2 {\hat{p}_\mu\hat{p}_\nu \over \hat{p}^2_d} 
    \widetilde{G}_{dd}^{(A)}({\bm p},\zeta_1) -\\
& \quad\qquad -r {\hat{p}_\mu\over \hat{p}_d} \widetilde{G}_{\nu d}^{(A)}({\bm p},\zeta_1)  -
    r {\hat{p}_\nu\over \hat{p}_d} \widetilde{G}_{\mu d}^{(A)}({\bm p},\zeta_1) \ ,\\ 
& r =
\left( 1 - \sqrt{\zeta_2\over \zeta_1}\right)\ . 
\end{aligned}
\end{equation}
To simplify this expression, we can use the Ward identity 
(see Sec.~\ref{sec:B}):
\begin{equation}
   \hat{p}_d \widetilde{G}_{d\mu}^{(A)}({\bm p},\zeta) = \zeta \hat{p}_\mu.
\label{Ward-axial}
\end{equation}
We end up with ($\mu,\nu\le d-1$)
\begin{equation}
\widetilde{G}_{\mu\nu}^{(A)}({\bm p},\zeta_2) = 
    \widetilde{G}_{\mu\nu}^{(A)}({\bm p},\zeta_1) + 
    (\zeta_2 - \zeta_1) {\hat{p}_\mu\hat{p}_\nu \over \hat{p}_d^2}.
\end{equation}
Taking the limit $\zeta_1\to 0$ his relation allows us to relate the hard-gauge
and soft-gauge susceptibilities. We find 
\begin{equation}
   \chi_{\mu\nu,\zeta} = \chi_{\mu\nu,HA} + \zeta,
\label{chiSA-HA}
\end{equation}
where the $\chi_{\mu\nu,\zeta}$ and $\chi_{\mu\nu,HA}$ are computed in the 
soft gauge with parameter $\zeta$ and in the hard gauge, respectively.

\subsection{Hard and soft Lorenz gauges}

The same calculation can be performed in the Lorenz case. We consider
\begin{equation}
\widetilde{\phi}({\bm p}) = 
\left( 1 - \sqrt{\zeta_2\over \zeta_1}\right)
   {1\over \hat{p}^2} \sum_{\mu} i \hat{p}_\mu
   \widetilde{A}_\mu({\bm p}),
\end{equation}
that allows us to map the model with parameter $\zeta_1$ onto the model with
parameter $\zeta_2$. Here $\hat{p}^2 = \sum_\mu \hat{p}^2_\mu$. The calculation
is analogous to that performed before. If we parametrize the susceptibilities
as in Eq.~(\ref{Lorenz-parametrization}), we obtain
\begin{eqnarray}
c_{L1}(\zeta_2) &=& 
    {d-1\over d}[c_{L1}(\zeta_1) - c_{L2}(\zeta_1)]
\nonumber \\
&&  + {\zeta_2\over d\zeta_1} [c_{L1}(\zeta_1) + (d-1) c_{L2}(\zeta_1)] ,
\nonumber \\
c_{L2}(\zeta_2) &=& 
    -{1\over d}[c_{L1}(\zeta_1) - c_{L2}(\zeta_1)] 
\nonumber \\ 
&& + {\zeta_2\over d\zeta_1} [c_{L1}(\zeta_1) + (d-1) c_{L2}(\zeta_1)] .
\label{chi-LSoft}
\end{eqnarray}
To simplify this expression, we use the Ward identity (see App.~\ref{sec:B})
\begin{equation}
    \sum_\mu \hat{p}_\mu \widetilde{G}_{\mu\nu}^{(L)}({\bm p},\zeta) = 
     \zeta {\hat{p}_\nu \over \hat{p}^2} , 
\label{Ward-Lorenz}
\end{equation} 
which implies 
\begin{equation} 
    \sum_\mu \widetilde{G}_{\mu\nu}^{(L)}({\bm p}_a,\zeta) = 
       {\zeta \over d \hat{p}^2_{\rm min}},
\end{equation}
with $p_{\rm min} = \pi/L$. Substituting in Eq.~(\ref{chi-LSoft}) we obtain 
\begin{eqnarray}
   c_{L1}(\zeta_2) &=&  c_{L1}(\zeta_1)  + 
   {1\over d^2 \hat{p}_{\rm min}^2 } (\zeta_2 - \zeta_1), \nonumber  \\
   c_{L2}(\zeta_2) &=&  c_{L2}(\zeta_1)  + 
   {1\over d^2 \hat{p}_{\rm min}^2 } (\zeta_2 - \zeta_1). 
\label{chiSL-chiHL}
\end{eqnarray}

\section{Numerical results}\label{sec:res}

To understand the role that the different gauge fixings play, we now discuss
the behavior of the gauge correlations in the three-dimensional Abelian-Higgs
(AH) model.  This lattice model has been extensively studied
\cite{Bonati:2020jlm, MV-08, KMPST-08} and we will use it as a
paradigmatic system to investigate how gauge correlations vary with the gauge
fixing adopted.

We consider $N$-dimensional scalar fields ${\bm z}_{\bm x}$, which are  defined
on the lattice sites and satisfy the unit-length constraint $\overline{\bm z}
\cdot {\bm z} = 1$. The matter action is
\begin{equation}
S_{\rm matter} = - J N \sum_{{\bm x},\mu} \hbox{Re }
({\overline{\bm z}}_{\bm x} \cdot \lambda_{{\bm x},\mu} 
     z_{{\bm x} + \hat{\mu}}),
\end{equation}
where the sum extends to all lattice sites and directions ($\mu$ runs 
from 1 to $d=3$), and 
$\lambda_{{\bm x},\mu} = \exp (i A_{{\bm x},\mu})$. 

\begin{figure}[tbp]
\includegraphics*[width=0.9\columnwidth]{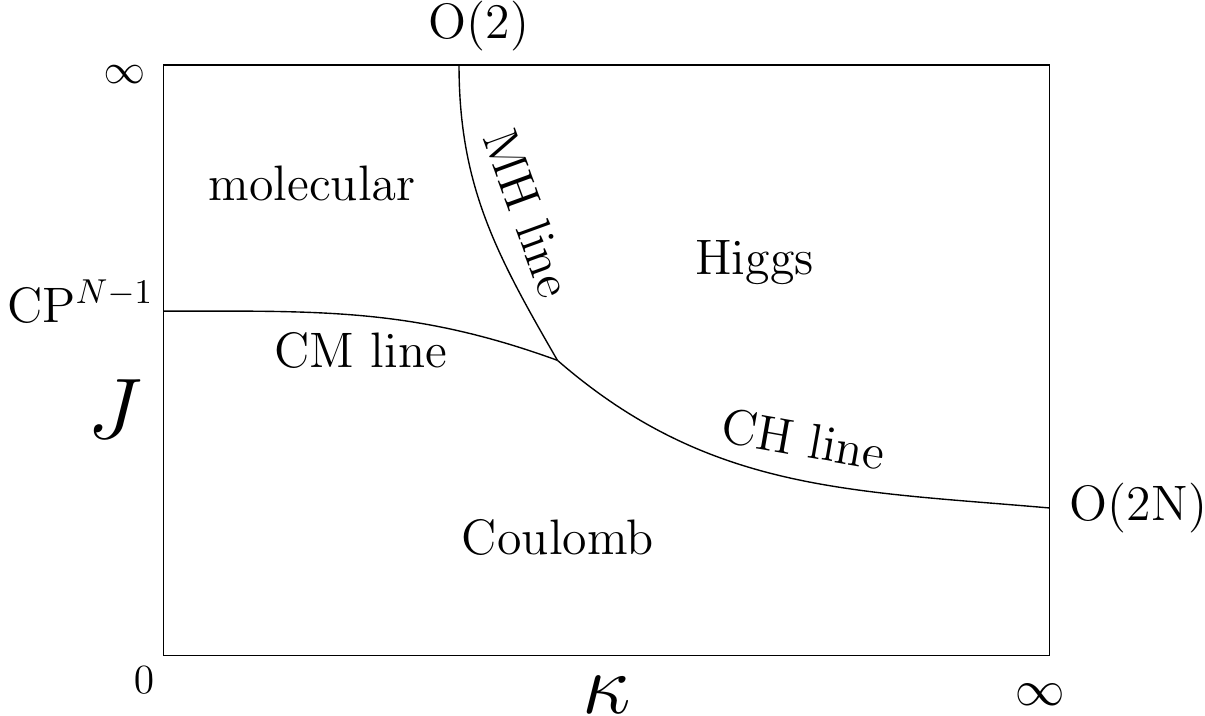}
  \caption{Sketch of the phase diagram of the three-dimensional 
    lattice AH model with
    noncompact gauge fields and unit-length $N$-component complex
    scalar fields, for generic $N\ge 2$.  Three transition lines can
    be identified: the Coulomb-to-Higgs (CH) line between the Coulomb
    and Higgs phases, the Coulomb-to-molecular (CM) line, and the
    molecular-to-Higgs (MH) line.  
    For $\kappa=0$, the model is equivalent to the
    CP$^{N-1}$ model, for $\kappa\to\infty$ to the O($2N$) vector,
    and for $J\to\infty$ to the inverted XY or O(2) model.}
\label{phdiasketch}
\end{figure}

The phase diagram is reported in Fig.~\ref{phdiasketch}.  It displays three
different phases characterized by the different behavior of the gauge field and
by the possible breaking of the global SU($N$) symmetry. For small $J$-values
the gauge field is expected to have long-range correlations as it occurs for
$J=0$ and the SU($N$) symmetry is realized in the spectrum (Coulomb phase). For
large $J$ two phases occur: the SU($N$) symmetry is broken in both phases,
while the gauge field is expected to be long-ranged for small $\kappa$
(molecular phase) and short-ranged for large $\kappa$ (Higgs phase).  The
properties of the Higgs phase are supposedly those that are usually associated,
in the perturbative setting, with the spontaneous breaking of the U(1) gauge
symmetry. The transition line separating the Coulomb and the Higgs phases is
the one along which (for $N\gtrsim 7$) the continuum limit associated with the
AH QFT emerges, while the other two transition lines are associated with  more
conventional critical behaviors, see Ref.~\cite{Bonati:2020jlm} for more details. 

In this work we consider scalar fields with $N=25$ components focusing on the
large-size behavior of the gauge observables in the Higgs and Coulomb phases.
We perform simulations for $(\kappa,J) = (0.4,0.2)$ and $(0.4,0.4)$  that lie
in the Coulomb and Higgs phase, respectively (for $\kappa = 0.4$, the
transition between the Coulomb and Higgs phases occurs \cite{Bonati:2021vvs} at
$J = 0.295515(4)$).  We report results for four different gauge fixings.  We
consider the hard Lorenz and axial gauge fixings and the corresponding soft
versions with $\zeta = 1$. We show that the long-distance behavior of the gauge
observables defined before depends, to some extent, on the gauge fixing used.
For the Coulomb case, the results are consistent with the ones that can be
analytically obtained for $J=0$, i.e., the noncompact Abelian lattice gauge
theory without matter, which are summarized in App.~\ref{sec:A}.

Simulations have been performed by using the same combination of Metropolis and
microcanonical updates discussed in Ref.~\cite{Bonati:2020jlm}, which can be
easily extended to the case of the soft gauges discussed in this paper.
Hard-axial simulations have been carried out by fixing $A_{{\bm x}, d}=0$ and
updating only the $d-1$ nonvanishing components of $A_{{\bm x},\mu}$.  To
obtain the results in the hard Lorenz gauge, we have instead performed
simulations with no gauge fixing and implemented the gauge fixing before each
measure. Given the gauge configuration $\{A_{{\bm x},\mu}\}$ obtained in the
simulation, we have determined a gauge transformation (\ref{eq:gaugetrans}) so
that the fields $\{A_{{\bm x},\mu}'\}$ satisfy the condition $F_{L,{\bm
x}}(A')=0$ for all $\bm x$ (see Eq.~\eqref{eq:Lorenzgauge}). Gauge correlations
are then computed using the fields $\{A_{{\bm x},\mu}'\}$.  The gauge
transformation has been determined by using a conjugate-gradient solver.

\subsection{Coulomb phase} 

\begin{figure}
\includegraphics*[width=0.95\columnwidth]{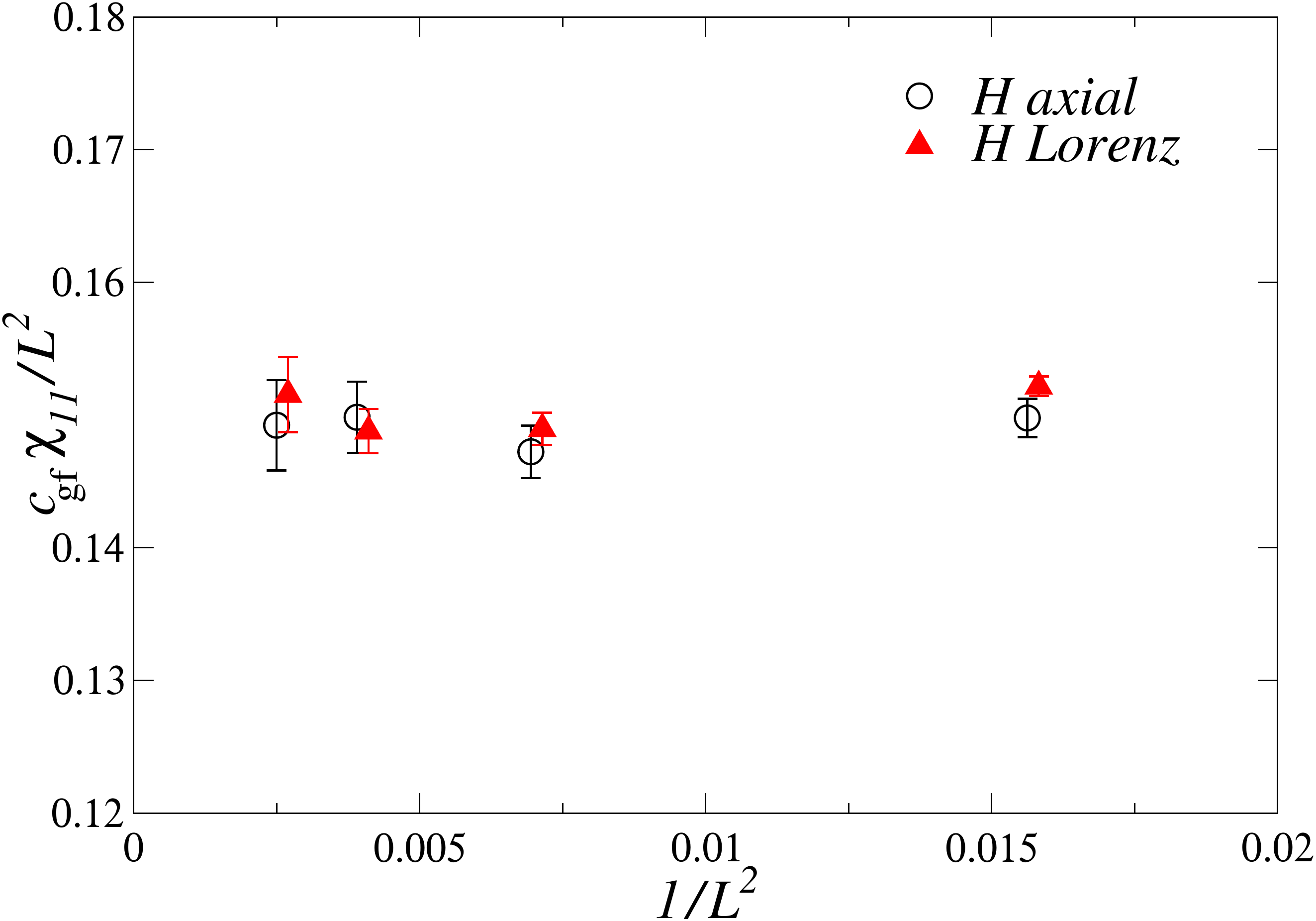}
\caption{
(Coulomb phase) Estimates of
$c_{\rm gf} \chi_{11}/L^2$ versus $1/L^2$ in the hard Lorenz and hard axial gauge,
where $c_{\rm gf}$ is a gauge-fixing dependent constant. We use $c_{\rm gf} = 3$ for
the Lorenz gauge fixing and $c_{\rm gf} = 1$ for the axial one. Lorenz gauge data
have been slightly shifted toward the right to improve readability.
Results in the Coulomb phase, for $J = 0.2$.  
}
\label{fig:Coulomb_chi}
\end{figure}

\begin{figure}
\includegraphics*[width=0.95\columnwidth]{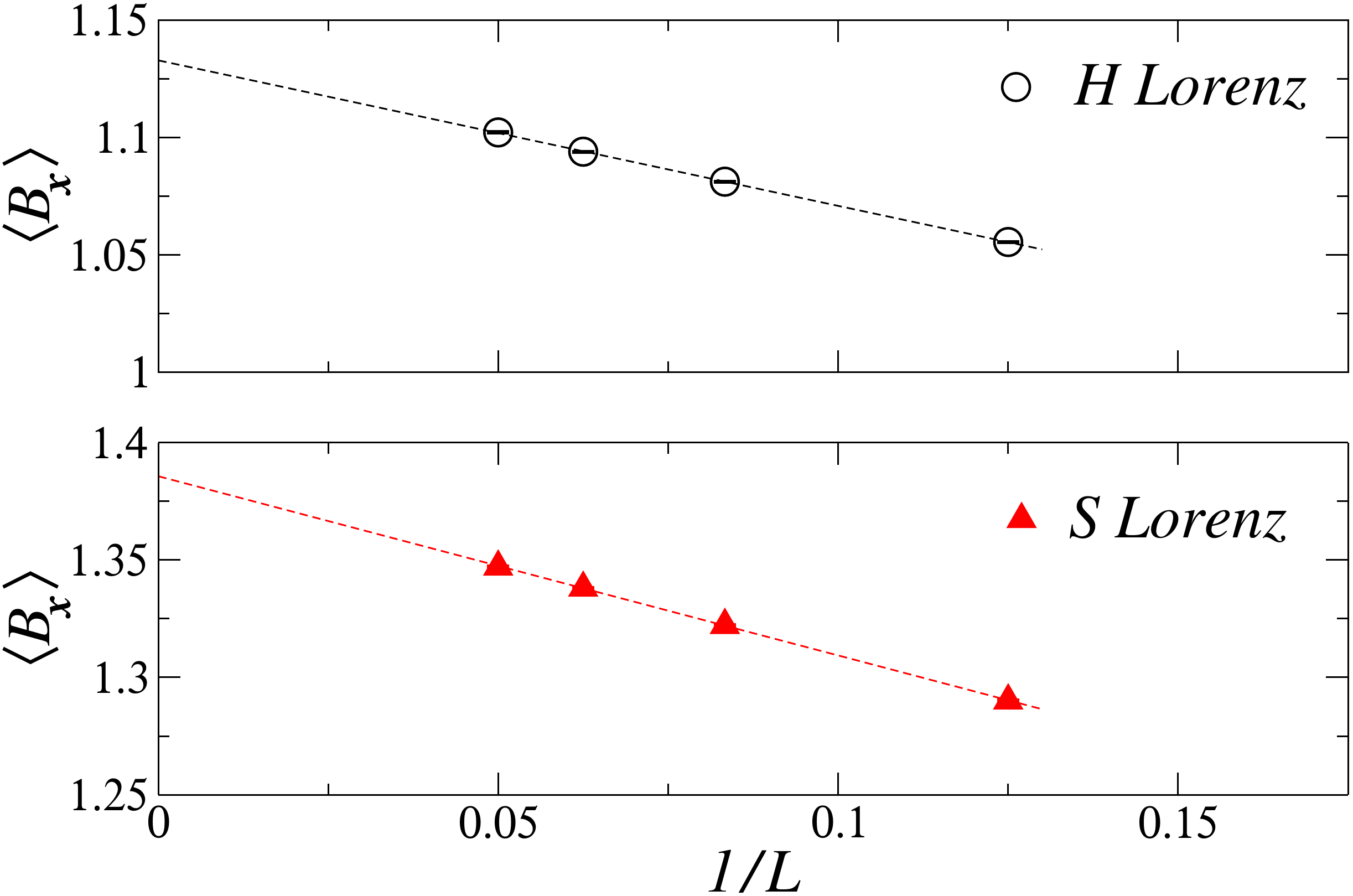}
\caption{
(Coulomb phase) Estimates of  $\langle B_{\bm x} \rangle$ versus $1/L$ in
the hard Lorenz (top) and soft Lorenz gauge with $\zeta=1$ (bottom). For $L\to\infty$
$\langle B_{\bm x} \rangle  \approx 1.1328$ and $1.3857$ in the two cases, respectively. 
Results in the Coulomb phase, for $J = 0.2$.
}
\label{fig:Coulomb_B_L}
\end{figure}

We start by investigating the behavior of the gauge model in the Coulomb phase
(simulations for $J=0.2$).  In the whole Coulomb phase the gauge field is
expected to have long-range correlations, and thus $\chi_{\mu\nu}$ should
diverge as $L$ increases, in all gauges considered. Results for the two hard
gauges are reported in Fig.~\ref{fig:Coulomb_chi}. We observe that
$\chi_{\mu\nu}$ diverges as $L^2$ in both cases, a fact that is consistent with the analytic
results for $J=0$ (in which case $1/L^2$ corrections are expected), see
App.~\ref{sec:A}.
The relation Eq.~\eqref{chiHA-chiHL} is fully confirmed
by the data, see Fig.~\ref{fig:Coulomb_chi}, and results in the soft gauges behave
analogously and are in full agreement with relations (\ref{chiSL-chiHL}) and
(\ref{chiSA-HA}).

\begin{figure}
\includegraphics*[width=0.95\columnwidth]{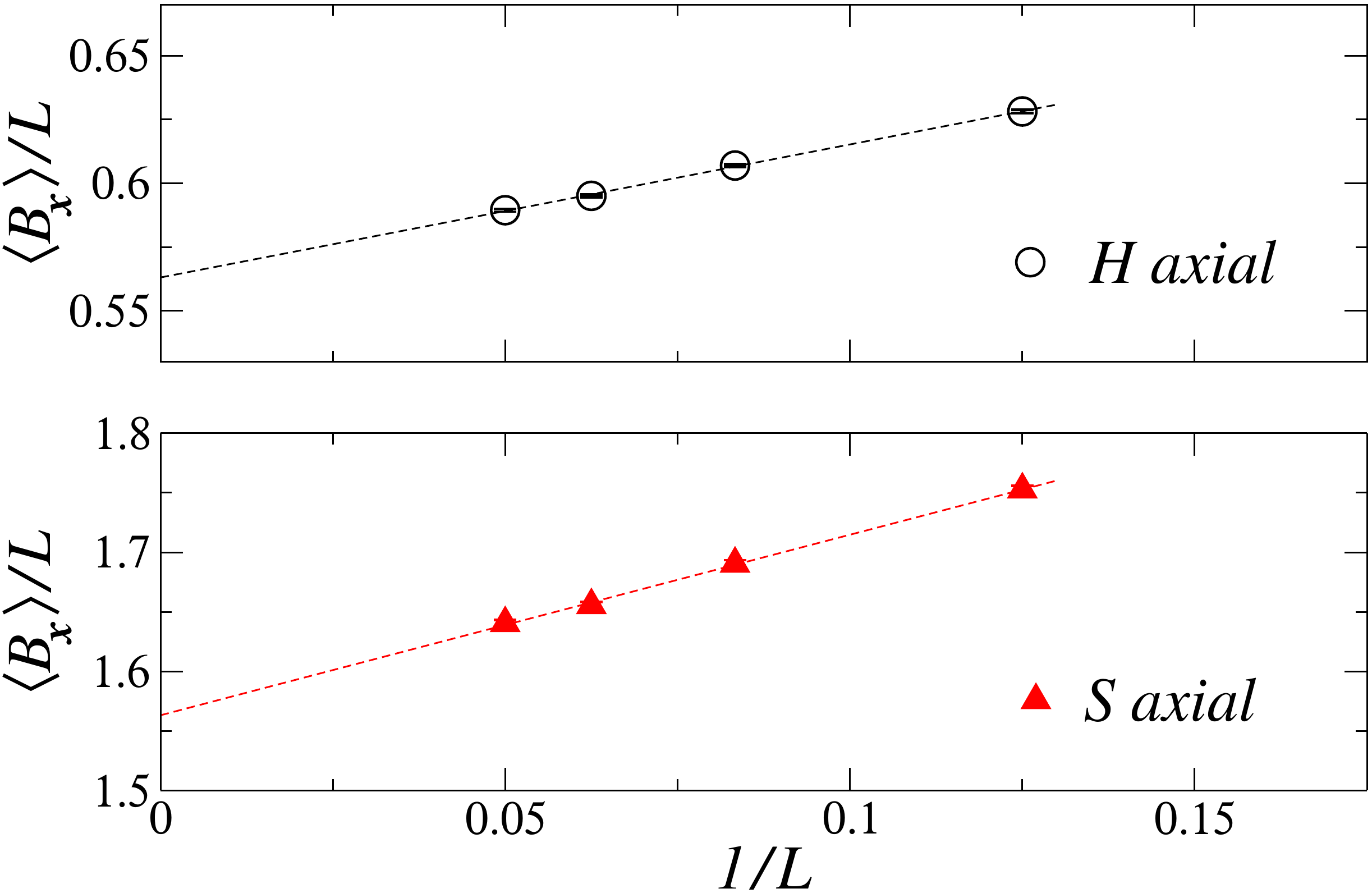}
\caption{
(Coulomb phase) Estimates of $\langle B_{\bm x} \rangle/L$ versus $1/L$ in
the hard axial (top) and soft axial gauge with $\zeta=1$ (botton). For $L\to\infty$
$\langle B_{\bm x}\rangle/L  \approx 
0.5631$ and $1.563$ in the two cases, respectively. 
Results in the Coulomb phase, for $J = 0.2$.
}
\label{fig:Coulomb_B_A}
\end{figure}

Let us now consider the average of the photon mass operator $B_{\bm x}$.
Results in the Lorenz gauges are reported in Fig.~\ref{fig:Coulomb_B_L}. 
In both case $\langle B_{\bm x}\rangle$ has a finite infinite-volume limit
with corrections of order $1/L$. Again this is in agreement with the 
results for $J=0$ reported in App.~\ref{sec:A}. We have determined the same 
quantity in the axial gauges obtaining a different result. In this case
$\langle B_{\bm x}\rangle$ diverges with the system size as $L$ increases,
see Fig.~\ref{fig:Coulomb_B_A}:  $B_{\bm x}$ is not a well-defined operator
in the infinite-volume limit. The different behavior can be understood by 
noting the completely different role the two gauge fixings play in 
infinite volume. In infinite volume, only the transformations 
$A_{{\bm x},\mu}' = A_{{\bm x},\mu} + c_\mu$, where $c_\mu$ is a constant, 
leave the Lorenz gauge-fixed action invariant. Indeed, in the Lorenz 
gauge, a gauge transformation leaves $A_{{\bm x},\mu}$ invariant only if 
\begin{equation}
\sum_{\mu} [\phi_{{\bm x} +\hat{\mu}} - 2 \phi_{\bm x} + 
            \phi_{{\bm x} -\hat{\mu}} ] = 0
\end{equation}
for all points ${\bm x}$.  By working in Fourier space, one can show that all
solutions of this equation can be written as $\phi_{\bm x} = a + \sum_\mu c_\mu
x_\mu$, so that, $\Delta_\mu \phi_{\bm x} = c_\mu$. Note that these 
gauge transformations are valid only in infinite volume.
In a finite volume with $C^*$ boundary conditions, the gauge fixing is 
complete and $c_\mu$ necessarily vanishes.

On the other hand, in the axial gauge, any gauge transformation with function
$\phi_{\bm x} = \phi(x,y,z)$ that only depends on $x$ and $y$ leaves the
action invariant.  Thus, the axial-gauge action is invariant under a large set
of space-dependent transformations and this causes the divergence of $\langle
B_{\bm x}\rangle$.  This result can also be understood by looking at the
relation between the axial and Lorenz correlation functions, see
Eq.~(\ref{GA-GL}).  While the Lorenz correlation function is expected to be
singular only for $p=0$ (due to the presence of the zero modes discussed
above), the axial correlation function is singular for $p_d = 0$, irrespective
of the value of the other components of the momentum, i.e., on a
$(d-1)$-dimensional momentum surface. These singularities make the sum
appearing in Eq.~(\ref{B-sum}) diverge as $L\to \infty$. 

It is well known that perturbation theory in the axial gauges is problematic
\cite{ZJ_book, Leibbrandt:1987qv}. The results presented here show that the difficulties one
encounters using axial gauges are not simply technical ones due to the infrared
problems of the perturbative expansion. Also nonperturbatively, axial gauges do
not allow a proper definition of some gauge-dependent quantities, for instance,
the photon-mass operator, in the infinite-volume limit.

We have also determined the behavior of the susceptibility 
$\chi_B$, obtaining results that are analogous to those that hold for 
$J=0$. We find $\chi_B\sim L$ in Lorenz gauges and  $\chi_B\sim L^3$ in 
axial gauges. 

Finally, let us make a few comments on
the apparently equivalent Lorenz gauge fixing
\begin{equation} 
F_{L',{\bm x}}(A) = \sum_{\mu=1}^d (A_{{\bm x}+\hat{\mu},\mu}-A_{{\bm x},\mu}),
\label{eq:Lorenzgauge2}
\end{equation}
which differs from the one reported in Eq.~(\ref{eq:Lorenzgauge}) in the choice
of the lattice derivative (forward instead of backward). This gauge fixing
function has several shortcomings. First of all, in even dimension, it does not
represent a complete gauge fixing for some values of $L$. For instance, if $L =
4 n + 2$ and $d=4$, transformations with [$\bm x=(x_1,x_2,x_3,x_4)$]
\begin{equation}
\phi_{\bm x} = A \cos \Bigl [{\pi\over 2} (x_1-x_2)\Bigr]
                          \cos \Bigl [{\pi\over 2} (x_3-x_4)\Bigr] 
\label{L2-example}
\end{equation}
leave $F_{L',{\bm x}}(A)$ invariant and are consistent with the $C^*$ boundary
conditions ($\phi_{\bm x}$ is antiperiodic). In $d=3$ the gauge fixing is
complete in a finite volume. However, in infinite volume, $F_{L',{\bm x}}(A)$
is invariant under a large set of gauge transformations, as it occurs in the
axial case. Thus, we do not expect $B_{\bm x}$ to be a well-defined operator if
$F_{L',{\bm x}}(A)$ is used. For $J=0$, the average value of $B_{\bm x}$
diverges in the infinite-volume limit, see App.~\ref{sec:A}.  We have also
performed some simulations for $J=0.2$, observing that also in this case
$\langle B_{\bm x}\rangle $ increases as $L$ is varied.

\subsection{Higgs phase}

\begin{figure}
\includegraphics*[width=0.95\columnwidth]{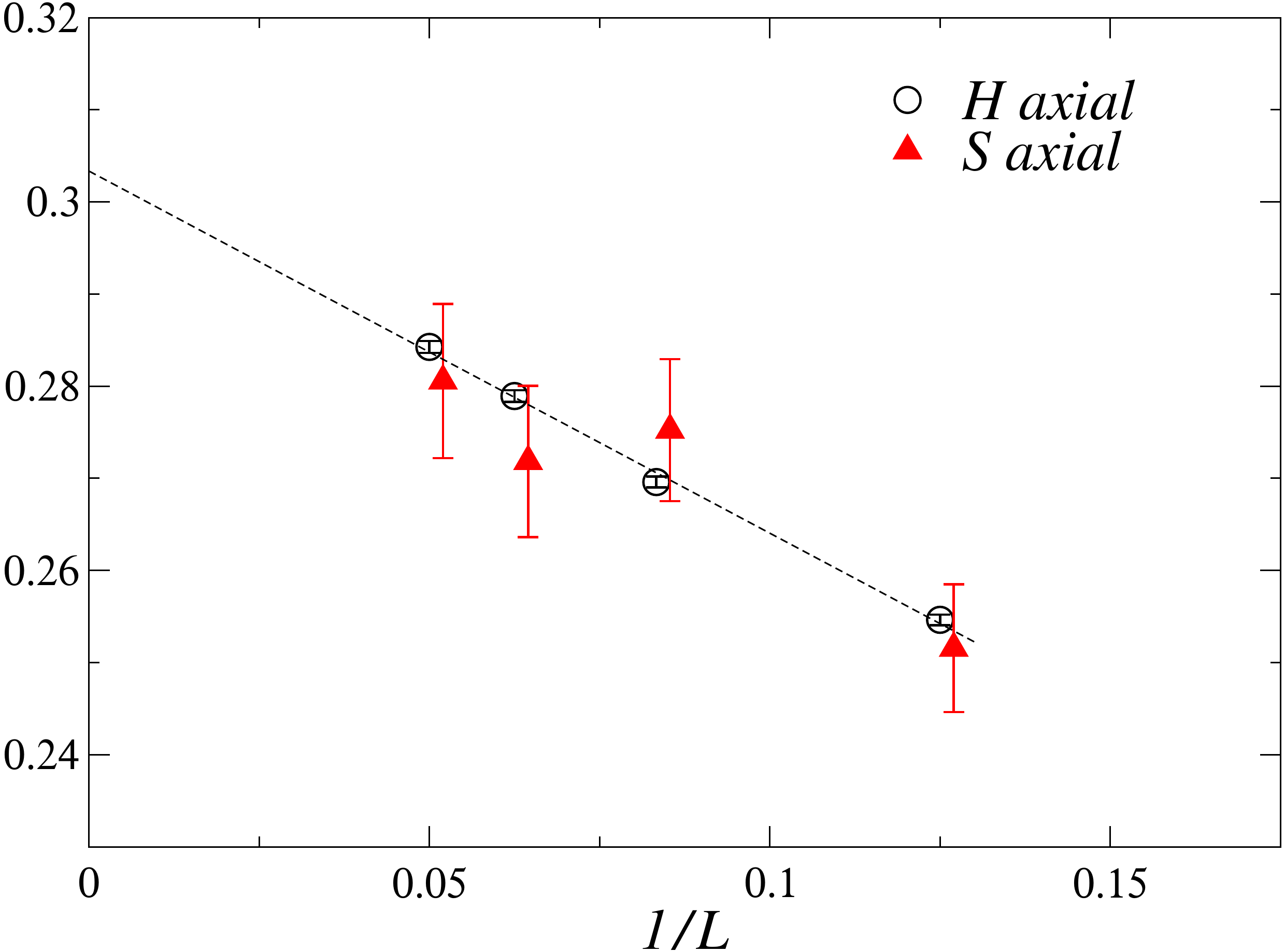}
\caption{
(Higgs phase) Estimates of  $\chi_{11}$ (hard axial gauge) and $\chi_{11} - 1$
(soft axial gauge with $\zeta = 1$), versus $1/L$, in the Higgs phase, $J=0.4$.
Soft axial gauge data have been slightly moved to the right to improve
readability. The line is only meant to guide the eye, since we have no 
theoretical understanding of the possibile origin of the $1/L$ correction.
}
\label{fig:Higgs-chi-A}
\end{figure}

\begin{figure}
\includegraphics*[width=0.95\columnwidth]{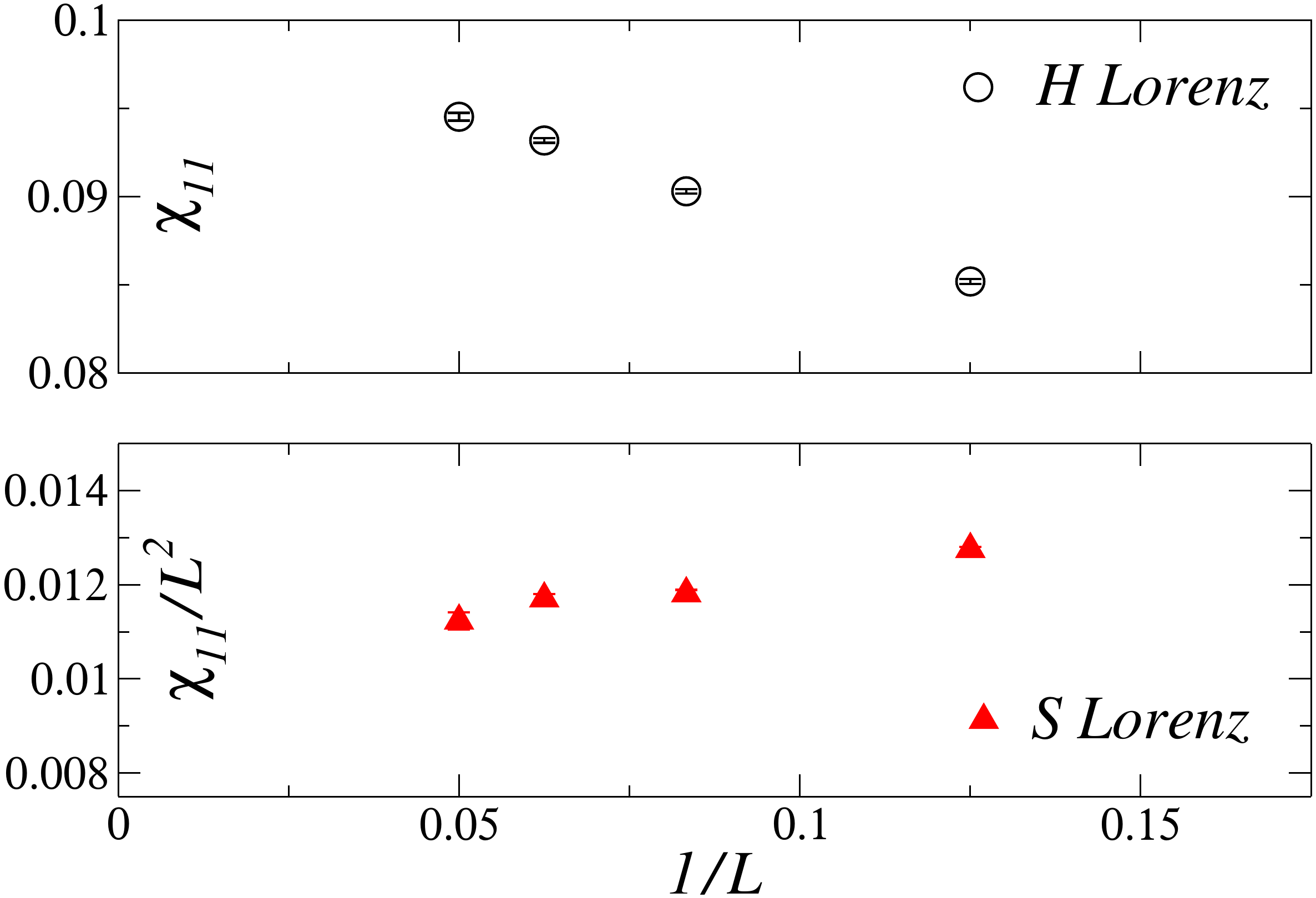}
\caption{
(Higgs phase) Top: Estimates of  $\chi_{11}$ in the hard Lorenz gauge; bottom:
Estimates of  $\chi_{11}/L^2$ in the soft Lorenz gauge with $\zeta=1$.  Results
in the Higgs phase for $J=0.4$.
}
\label{fig:Higgs-chi-L}
\end{figure}

\begin{figure}
\includegraphics*[width=0.95\columnwidth]{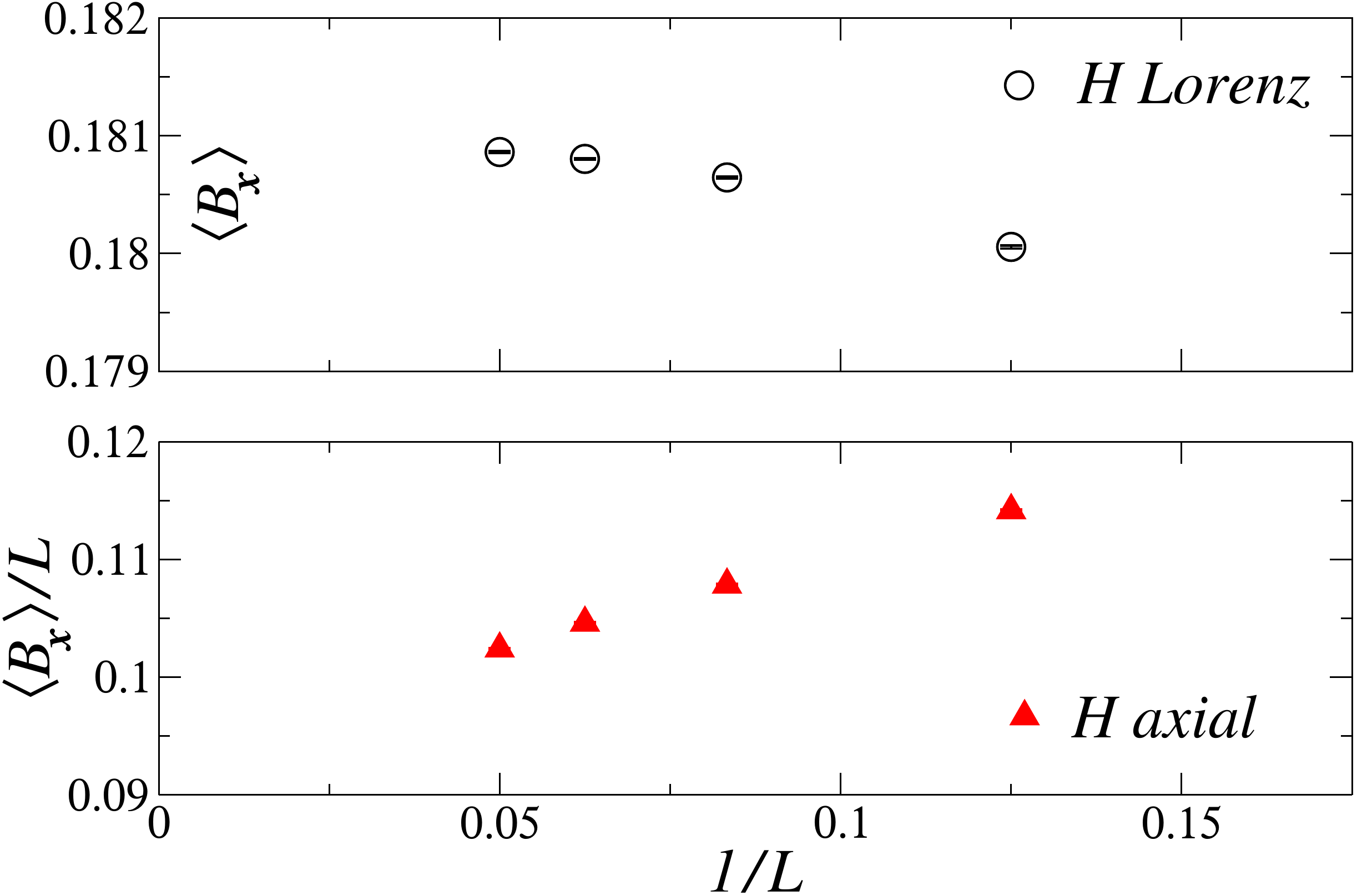}
\caption{
(Higgs phase) Top: Estimates of  $\langle B_{\bm x} \rangle$ in the hard Lorenz
gauge; bottom: Estimates of  $\langle B_{\bm x}\rangle /L $ in the hard axial
gauge.  Results in the Higgs phase for $J=0.4$.
}
\label{fig:Higgs-B}
\end{figure}

Let us now discuss the behavior of gauge-dependent observables in the Higgs
phase (numerical simulations have been performed for $J=0.4$).  In
Fig.~\ref{fig:Higgs-chi-A} we report the susceptibility $\chi_{11}$ for the
hard and the soft axial gauge (with $\zeta = 1$), versus $1/L$. In both cases
$\chi_{11}$ has a finite limit as $L\to \infty$ and satisfies relation
(\ref{chiSA-HA}).  The finite value in the Higgs phase is consistent with the
presence of a finite photon mass.  However, the apparent presence of size
corrections that decay as $1/L$ points to an unusual behavior of the system,
since in a standard massive phase corrections are typically expected to scale
as $e^{-L/\xi}$. 

In Fig.~\ref{fig:Higgs-chi-L} we show results for the susceptibility
$\chi_{11}$ in the Lorenz gauges. In the hard case, $\chi_{11}$ is finite in
the infinite-volume limit and satisfies the exact relation (\ref{chiHA-chiHL})
with the corresponding quantity in the hard axial gauge. Instead, in the soft
Lorenz gauge, we find $\chi_{11} \sim L^2$. This divergence might be,
erroneously, interpreted as an indication of the presence of {\em physical}
long-range gauge correlations in the Higgs phase---this would be in contrast
with the idea that the photon is massive.  The correct interpretation is
instead, that in the soft Lorenz gauge that are {\em unphysical} gauge modes
that are long-ranged and contribute to $\chi_{\mu\nu}$, even though they do not
have physical meaning.  This interpretation is supported by
Eq.~(\ref{chiSL-chiHL}) that we rewrite as 
\begin{equation}
\chi_{11,L\zeta} = \chi_{11,HL} + {\zeta \over d^2 \hat{p}_{\rm min}^2} 
\approx \chi_{11,HL} + {\zeta \over d^2\pi^2} L^2
\end{equation}
where $\chi_{11,L\zeta}$ and $\chi_{11,HL}$ refer to the soft Lorenz 
gauge with parameter $\zeta$ and to the hard Lorenz gauge, respectively. 
Since $\chi_{11,HL}$ has a finite large-$L$ limit,
this relation shows that the divergence of $\chi_{11,L\zeta}$ 
is only due to the last term, which has no physical meaning, and is related 
to the presence of propagating longitudinal modes that are instead completely
suppressed in the hard gauge ($\zeta = 0$).

Perturbation theory provides the recipe for the definition of a susceptibility
that only couples the physical modes. We define 
\begin{equation} \label{eq:Gtr}
   \widetilde{G}_{\rm tr}({\bm p}) = \sum_{\mu\nu} 
    \left(1 - {\hat{p}_\mu \hat{p}_\nu\over \hat{p}^2} \right) 
     \widetilde{G}_{\mu\nu}({\bm p}).
\end{equation}
and a transverse susceptibility 
$\chi_{\rm tr} = \widetilde{G}_{\rm tr}({\bm p}_a)$. 
Using the parametrization (\ref{Lorenz-parametrization}) we obtain
\begin{equation}
   \chi_{\rm tr}  = (d-1) (c_{L1} - c_{L2}).
\end{equation}
Eq.~(\ref{chiSL-chiHL}) then implies 
\begin{equation} 
   \chi_{\rm tr}(\zeta_1) = \chi_{\rm tr}(\zeta_2).
\end{equation}
The transverse susceptibility is independent of $\zeta$ and therefore is the
same in hard and soft gauges. In particular, $\chi_{\rm tr}(\zeta)$ is finite
in the Higgs phase for all values of $\zeta$, as expected.

The behavior of $\langle B_{\bm x}\rangle$ and of the corresponding
susceptibility is analogous to that observed in the Coulomb phase, see
Fig.~\ref{fig:Higgs-B}.  The average $\langle B_{\bm x}\rangle$ is well defined
only for Lorenz gauges. In the axial gauge, we have instead $\langle B_{\bm
x}\rangle \sim L$.  This is not unexpected since the argument we have presented
in the previous section, i.e., that the divergence of $\langle B_{\bm
x}\rangle$ is related to the large number of quasi-zero modes present in the
axial case, does not rely on any particular property of the two phases.

Finally, let us consider the susceptibility $\chi_B$. Not surprisingly, in the
axial gauge data are consistent with a behavior $\chi_B\sim L^3$, as in the
Coulomb phase. In the hard Lorenz case, we observe that $\chi_B$ is finite as
$L$ increases. This is the expected behavior in the Higgs phase, in which the
photon is massive. In the soft Lorenz gauge instead, data are consistent with
$\chi_B\sim L$. It is easy to realize that this divergence is due to the
contributions of the nonphysical longitudinal modes present for nonzero values
of $\zeta$. The linear divergence with $L$ can be predicted by a simple
argument. Let us assume that the hard-gauge correlation function has
the form (at least for small values of $\bm p$)
\begin{equation}
   \widetilde{G}_{\mu\nu} ({\bm p},\zeta=0) = 
   {Z \over \hat{p}^2 + M^2} \left( \delta_{\mu\nu} - 
    { \hat{p}_\mu \hat{p}_\nu\over \hat{p}^2} \right),
\end{equation}
and, as predicted by the Ward identities, that 
\begin{equation}
   \widetilde{G}_{\mu\nu} ({\bm p},\zeta) = 
     \widetilde{G}_{\mu\nu} ({\bm p},\zeta=0) + 
    \zeta { \hat{p}_\mu \hat{p}_\nu\over (\hat{p}^2)^2 }.
\end{equation}
In a Gaussian approximation---we neglect irreducible four-field
contributions---we have 
\begin{equation}
   \chi_B = {2\over L^d} \sum_{\bm p} \sum_{\mu\nu}
     \widetilde{G}_{\mu\nu} ({\bm p},\zeta) \widetilde{G}_{\mu\nu} (-{\bm p},\zeta),
\end{equation}
and therefore, 
\begin{equation}
\begin{aligned}
\chi_B &= 2 (d-1) Z^2 {1\over L^d} \sum_{\bm p} {1\over (\hat{p}^2 + M^2)^2} + \\
     &+2 \zeta {1\over L^d} \sum_{\bm p} {1\over (\hat{p}^2)^2}.
\end{aligned}
\end{equation}
The first sum has a finite limit as $L\to \infty$, while the second one,
see App.~\ref{sec:A}, diverges as $L$ and $\ln L$ in $d=3$ and $d=4$,
respectively. Thus, in three dimensions 
the longitudinal modes give rise to a contribution
that increases as $L$, in agreement with the numerical results.
We conclude that the photon mass operator is not well-defined
nonperturbatively in the Lorenz soft gauge, because of the contributions
of the nonphysical longitudinal modes. Apparently, only the hard Lorenz
gauge is a consistent gauge fixing in which the operator is correctly
defined.

\section{Some field theory results} \label{sec:FT}

The results of the previous sections can be combined with QFT
results to obtain some general predictions of the behavior of 
Abelian gauge systems at charged fixed points.

First, let us  note that our previous results also allow us to predict that the
anomalous dimension of the gauge field is the same in the axial gauge as in the
Lorenz gauge.  Indeed, as we have discussed before, the large-scale behavior of
the susceptibilities $\chi_{\mu\nu}$ (for $\mu,\nu<d$) is the same for all gauge
fixings (although some caution should be exercised in the soft Lorenz case).
Indeed, a summary of the results obtained is the following: 
\begin{enumerate}[i.]
\item the susceptibilities
$\chi_{\mu\nu}$ ($\mu,\nu < d$) 
in the hard Lorenz and in the hard axial gauge differ only by 
a multiplicative constant: $2d/(d-1)$ for $\mu=\nu$ and 
$-d$ for $\mu\not=\nu$, see Eq.~\eqref{chiHA-chiHL};
\item the susceptibilities in the hard and soft axial gauges 
differ by an additive constant, see Eq.~\eqref{chiSA-HA};
\item the susceptibilities in the hard and soft Lorenz gauge 
behave differently, because of the coupling with the longitudinal
modes. If one considers the transverse definition, see Eq.~\eqref{eq:Gtr},
results are independent of $\zeta$, i.e., are the same in the 
hard and soft case.
\end{enumerate}
For the soft Lorenz
gauge, one can prove to all orders of perturbation theory
that~\cite{HT96,ZJ_book} $\eta_A = 4-d$, independently of the nature of the
matter fields. Indeed, the proof only relies on the relation $Z_AZ_e=1$ between
the renormalization constants of the gauge field and of the electric charge
$e$. This implies \cite{ZJ_book}
\begin{equation}
\beta_{e^2}=\hat{e}_r^2(d-4+\eta_A)\ ,
\end{equation}
which connects the anomalous dimension $\eta_A$ of the gauge field, the $\beta$
function of the dimensionless charge $\hat{e}^2=e^2\mu^{d-4}$ ($\mu$ is the RG
scale), and the renormalized dimensionless charge $\hat{e}_r$.
At a transition which is associated with a charged fixed point, i.e., where the 
gauge theory provides the effective critical behavior, we have 
$\hat{e}_r^2 \not=0$. Therefore, the fixed-point condition $\beta_{e^2} = 0$
implies \cite{HT96}
\begin{equation}
\eta_A = 4 - d.
\label{etaA}
\end{equation}
Numerical results \cite{BPV-in-prep} for the 
three-dimensional Abelian-Higgs model are in full agreement with this 
prediction.

A second interesting result concerns the parameter $\zeta$ that parametrizes
the soft gauges. As a consequence of the Ward identities discussed in 
Sec.~\ref{sec:B}, in the soft Lorenz gauge we have $\zeta = \zeta_r Z_A$, which
implies 
\begin{equation}
\beta_{\zeta} = - \zeta_r \eta_A.
\label{betazeta}
\end{equation}
The value $\zeta = 0$ is a fixed point of this equation, as expected. Indeed,
if we start 
from a model with a purely transverse gauge field, no longitudinal contributions
are generated by the RG flow. Instead, if we start the flow from a value $\zeta \not=0$, 
$\zeta$ flows towards $+\infty$, indicating that the hard gauge fixing is an
unstable fixed point, at least for $d < 4$. 
Moreover, for $\zeta\not=0$ the large-scale behavior
is singular, as the nongauge-invariant modes become unbounded under the 
RG transformations. Therefore, also QFT (which describes the critical 
behavior at charged transitions) predicts that only the hard Lorenz gauge 
fixing provides a consistent definition of nongauge-invariant quantities at the 
critical point in three dimensions.

Eqs.~(\ref{etaA}) and (\ref{betazeta}) allow us to predict the crossover
behavior of $\chi_{\mu\nu}$ at a critical charged transition point in
the soft Lorenz gauges. For $d < 4$ we predict
\begin{equation}
   \chi_{\mu\nu}(\alpha) = L^{2-\eta_A} f_{\mu\nu} (\zeta L^{\eta_A}) = 
    L^{d-2} f_{\mu\nu} (\zeta L^{4-d}).
\end{equation}
This relation should hold for $L\to\infty$, $\zeta\to 0$ at fixed 
$\zeta L^{4-d}$. The function $f_{\mu\nu}(x)$ can be computed using 
Eq.~(\ref{chiSL-chiHL}). If 
$\chi_{\mu\nu}(\alpha=0) \approx a_{\mu\nu} L^{d-2}$
for $L\to \infty$, Eq.~(\ref{chiSL-chiHL}) implies 
\begin{equation}
\begin{aligned}
   \chi_{\mu\nu}(\alpha) & \approx a_{\mu\nu} L^{d-2} + 
    {\zeta L^2 \over d^2 \pi^2} =  \\
    &=L^{d-2} \left(a_{\mu\nu} + {1\over d^2\pi^2} \zeta L^{4-d} \right),
\end{aligned}
\end{equation}
so that $f_{\mu\nu}(x)= a_{\mu\nu} + x/(d^2 \pi^2)$.

\section{Conclusions}\label{sec:concl}

In this work we investigate the behavior of gauge correlations in Abelian gauge
theories with noncompact gauge fields. Because of the unbounded nature of the
fluctuations of the gauge fields, a rigorous definition of the model requires
the introduction of a gauge fixing term. This is at variance with compact
formulations (for instance, models with Wilson action), in which a gauge fixing
is not required to make the model well defined. Here we consider two widely
used gauge fixings, the axial and Lorenz one. We also distinguish between hard
gauge fixings---in this case the partition function is given in
Eq.~(\ref{Zhard})---and soft ones depending on a parameter $\zeta$---the
corresponding partition function is given in Eq.~(\ref{Zsoft}). 

Gauge-invariant correlations are obviously independent of the gauge-fixing
procedure. On the other hand, the large-scale behavior of gauge-dependent
quantities may have a nontrivial dependence. Here we first consider
correlations of the gauge field $A_{{\bm x},\mu}$ and we derive general
relations, independent of the nature of the matter couplings,  between these
correlations computed in the presence of different gauge fixings. Second, we
consider the photon-mass composite operator $A_{{\bm x},\mu}^2$, which is
usually introduced in the action, in perturbative calculations, as an infrared
regulator of the theory. 

As a specific example, we analyze the behavior of these correlation functions
in the three-dimensional Abelian-Higgs model, in which an $N$-component complex
scalar field is coupled with a noncompact real Abelian gauge field.  In
particular, we study their behavior in the so-called Coulomb and Higgs phases
(see Fig.~\ref{phdiasketch} for a sketch of the phase diagram). In the Coulomb
phase, the correlation function $\widetilde{G}_{\mu\nu}({\bm p})$ of the gauge
fields has the same 
small-momentum behavior as in the absence of matter fields, for
all gauge fixings considered. In particular, the susceptibiity $\chi_{\mu\nu}$
defined in Eq.~(\ref{def-chi}) diverges as $L^2$ in the infinite-volume limit.
In the Higgs phase, we expect the photon to be massive and therefore
$\chi_{\mu\nu}$ should be finite as $L\to \infty$.  This turns out to be true
for the axial soft and hard gauges and for the hard Lorenz gauge. On the other
hand, $\chi_{\mu\nu} \sim L^2$ in the soft Lorenz gauge.  This divergence is
caused by the unphysical contributions due to the longitudinal modes that
propagate in the soft Lorenz gauge. 

While the behavior of $\widetilde{G}_{\mu\nu}({\bm p})$ in all gauges is consistent
with the general picture that the photon is massless/massive in the
Coulomb/Higgs phase, the interpretation of the results for the photon mass operator $B_{\bm x} =
\sum_\mu A_{{\bm x},\mu}^2$ are more complicated. If we consider the soft
and hard axial gauges, we find $\langle B_{\bm x}\rangle \sim L$ in both
phases.  The operator does not have a well-defined infinite-volume limit. The
divergence is due to the presence of a $(d-1)$-dimensional family of quasi-zero
modes, so that $A_{{\bm x},\mu}$ develops infinite-range fluctuations in the
infinite-volume limit. Therefore, if an axial gauge fixing is used, $B_{\bm x}$
cannot be defined nonperturbatively.  In the soft and hard Lorenz gauge, the
average $\langle B_{\bm x}\rangle$ is finite as $L\to\infty$ in both phases,
and thus the operator is well defined. However, in the Higgs phase, the
susceptibility $\chi_B$ defined in Eq.~(\ref{def-chi}) behaves differently in
the hard and soft  case. In the hard case, $\chi_B$ has a finite
infinite-volume limit, as expected--- the photon mass is finite. Instead,
$\chi_B$ diverges as $L$ in the soft gauge. This divergence is due to the
longitudinal modes that are not fully suppressed. 

The results presented here show that neither the axial gauge nor the 
soft Lorenz gauge are appropriate for the study of generic gauge-dependent correlation
functions. The first type of gauges suffers from the existence of an infinite
family of quasi-zero modes, giving rise to spurious divergences, unrelated with 
the presence of long-range physical correlations. Soft Lorenz gauges suffer instead
from the presence of propagating unphysical longitudinal modes, that,
at least for $d < 4$ and therefore in three dimensions,  may hide 
the physical signal. Apparently, only the hard Lorenz gauge fixing provides 
a consistent model in which gauge-dependent correlations have
the expected large-scale (small-momentum) behavior.  It is interesting to
observe that also QFT singles out the hard Lorenz gauge as the 
gauge of choice for the study of gauge correlations. 
Note that the shortcomings
of the axial gauge and of the soft Lorenz gauge are not related to the 
nature of the matter fields but are due to intrinsic properties of 
the gauge fixings. Therefore, our conclusions should be relevant also 
for systems in which fermions are present.

\emph{Acknowledgement}. Numerical simulations have been performed on the CSN4
cluster of the Scientific Computing Center at INFN-PISA.

\appendix

\section{Critical behavior in the U(1) abelian gauge theory}\label{sec:A}

In this Appendix we summarize the expressions of the observables 
defined in Sec.~\ref{sec:2} for the free U(1) gauge theory, i.e., 
in the absence of matter fields.
The susceptibilities $\chi_{\mu\nu}$ can be 
trivially derived from the small-momentum behavior of 
$\widetilde{G}_{\mu\nu}({\bm p})$, defined in Sec.~\ref{sec:2}. 
Moreover, we have
\begin{eqnarray}
\langle B_{\bm x} \rangle &=& {1\over V} \sum_{{\bm p}} 
    \sum_{\mu\nu} \widetilde{G}_{\mu\nu} ({\bm p}),   \\
\chi_B &=& {2\over V} \sum_{{\bm p}}
    \sum_{\mu\nu} [\widetilde{G}_{\mu\nu} ({\bm p})]^2 \; .
\end{eqnarray}
Because of the $C^*$ boundary conditions the sums go over the momenta 
\begin{equation}
{\bm p} = {\pi \over L} \left (2 n_1 + 1, 2 n_2 + 1, 2 n_3 + 1\right),
\end{equation}
with $0 \le n_i < L$.

\subsection{Lorenz gauge}

In the Lorenz gauge, the propagator $\widetilde G_{\mu\nu}({\bm p})$ is given by
\begin{equation}
   \widetilde G_{\mu\nu}({\bm p}) = 
    {1\over \kappa} {\delta_{\mu\nu}\over \hat{p}^2} + 
   {\zeta\kappa - 1 \over \kappa} 
     {\hat{p}_\mu\hat{p}_\nu\over (\hat{p}^2)^2 },
\end{equation}
where $\hat{p}_\mu = 2 \sin p_\mu/2$ and
$\hat{p}^2 = \sum_\mu \hat{p}_\mu^2$. It follows that 
\begin{equation}
\begin{aligned}
 \chi_{\mu\nu} &= (d \delta_{\mu\nu} + \zeta \kappa - 1) 
   {1\over \kappa d^2 \hat{p}_{\rm min}^2}\ ,  \\
\langle B_{\bm x} \rangle &= {d-1 + \zeta\kappa\over \kappa} I_{d,1}(L)\ , \\
\chi_B &=  {2 (d-1 + \zeta^2 \kappa^2) \over \kappa^2} I_{d,2}(L)\ ,
\end{aligned}
\end{equation}
where $p_{\rm min} = \pi/L$ and 
\begin{equation}
   I_{d,n}(L) = {1\over L^d} \sum_{\bm p} {1\over (\hat{p}^2)^n}.
\end{equation}
The behavior of the sums $I_{d,n}$ depends on the dimension $d$. 
For $d > 2$, $I_{d,1}$ has a finite limit for $L\to \infty$, while 
it diverges logarithmically in $d=2$.
In particular, in $d=3$ we have 
\cite{GZ-77,JZ-01}
\begin{equation}
\begin{aligned}
& I_{3,1}(L)  \approx \int_{[-\pi,\pi]^3} {d^3p\over (2 \pi)^3} {1\over \hat{p}} =\\
& = 
  {1\over 192 \pi^3} (\sqrt{3} - 1) 
    \Gamma\left({1\over 24}\right)^2 
    \Gamma\left({11\over 24}\right)^2\approx\\
& \approx 0.252731\ .
\end{aligned}
\end{equation}
Instead, the sum $I_{d,2}$ diverges for $L\to \infty$ in dimension $d\le 4$, as
$L^{4-d}$ (as $\ln L$ in $d=4$).  We find 
\begin{equation}
\begin{aligned}
& I_{3,2}(L) \approx a_2 L [1 + O(L^{-1})] & & a_2 \approx 0.015216 \; , \\ 
& I_{4,2}(L) \approx a_2 \ln L + O(1) & & a_2 \approx {1\over 8 \pi^2} \; .
\end{aligned}
\end{equation}
Thus, in three dimensions the susceptibilities $\chi_{\mu\nu}$ 
and $\chi_B$ diverge as $L^2$ and $L$,
respectively, while $\langle B_{\bm x} \rangle$ is finite.

\subsection{Axial gauge}

In the axial gauge 
\begin{eqnarray}
\widetilde G_{\mu\nu}({\bm p}) &=&  {1\over \kappa}
  {\delta_{\mu\nu}\over \hat{p}^2} +
         {\hat{p}_\mu \hat{p}_\nu \over \hat{p}_d^2 } 
   \left({1\over \kappa \hat{p}^2} + \zeta \right),
\end{eqnarray}
if both $\mu$ and $\nu$ are not equal to $d$. Otherwise, we have 
\begin{eqnarray}
\widetilde G_{\mu\nu} ({\bm p}) &=& 
    \zeta {\hat{p}_\mu \hat{p}_\nu \over \hat{p}_d^2} . 
\end{eqnarray}
As for the susceptibilities, we find  $\chi_{d\mu} = \zeta$ and,
for $\mu,\nu< d$, 
\begin{equation}
\chi_{\mu\nu}
{\delta_{\mu\nu} + 1 \over d\kappa} {1\over \hat{p}_{\rm min}^2} + \zeta
  \approx  {\delta_{\mu\nu} + 1 \over d\kappa \pi^2} L^2 , 
\end{equation}
where $p_{\rm min} = \pi/L$. As expected, 
$\chi_{d\mu}$ is finite (it vanishes in the 
hard axial gauge for which $\zeta = 0$),
while the other susceptibility components diverge as $L^2$. Although 
the large-$L$ behavior is the same  as in the Lorenz case, here 
the asymptotic behavior is $\zeta$ independent: the susceptibilities
behave identically in the hard and soft axial case, a result that does not
hold in the Lorenz case.

As for $\langle B_{\bm x} \rangle$ and $\chi_B$
we find 
\begin{equation}
\begin{aligned}
& \langle B_{\bm x} \rangle = {1\over \kappa} ((d-2) I_{d,1} + J_1) + 
   \zeta (1 + (d-1)  J_1 J_{-1}),  \\
& \chi_B = 2 \zeta^2+ \\
& + 2 \zeta^2 (d-1)
   (2 J_1 J_{-1} + (d-2) J_2 J_{-1}^2 + J_2 J_{-2})  \\ 
& + {2\over \kappa^2} 
 ((d-2) I_{d,2} + J_2 + 2 (d-1) \zeta \kappa J_2 J_{-1}),
\end{aligned}
\end{equation}
where the quantities $J_n(L)$ correspond to the one-dimensional sums 
($p = (2 n + 1)\pi/L$ with $n=0,\dots,L-1$)
\begin{equation}
   J_n(L) = {1\over L} \sum_{p} \hat{p}^{-2n}.
\end{equation}
Since we have (these expressions can be derived as in App.~B.1.d
of Ref.~\cite{CP-98}) 
\begin{equation}
\begin{aligned}
   J_2 &= {1\over 48} L (L^2 + 2), \\
   J_1 &= {L\over 4}, \\
   J_{-1} &= 2, \\
   J_{-2} &= 6,
\end{aligned}
\end{equation}
we obtain for large values of $L$ for $d > 2$:
\begin{equation}
\begin{aligned}
\langle B_{\bm x} \rangle &\approx 
     {1 + 2(d-1) \zeta \kappa \over 4 \kappa} L,  \\
\chi_B  &\approx {1\over 24\kappa^2} [1 + 4(d-1) \zeta \kappa +   \\
   & \quad +2 (d-1) (2 d-1) \zeta^2 \kappa^2] L^3 .
\end{aligned}
\end{equation}
Note that $\langle B_{\bm x} \rangle$ diverges, at variance with what 
happens in the Lorenz case. From a technical point of view this is due to the
fact that the axial-gauge propagator is more divergent than the Lorenz one: 
indeed, in the axial gauge $\widetilde{G}_{\mu\nu}({\bm p})$ diverges as $p_d \to 0$,
for any value of the other momentum components, 
while in the Lorenz gauge a divergence 
is only observed as $|{\bm p}| \to 0$. 
More intuitively, note that, in infinite   
volume, the axial-gauge fixed Hamiltonian is still invariant under the
gauge transformations (\ref{eq:gaugetrans}) if the function  
$\phi_{\bm x}= \phi_{(x_1,\ldots,x_d)}$ depends on 
$x_i$ with $i < d$ only. This should be compared with the Lorenz case, in 
which only gauge transformations with $\Delta_\mu \phi_{\bm x} = c_\mu$, 
where $c_\mu$ is $x$-independent, 
leave the infinite-volume gauge-fixed Hamiltonian invariant.
The presence of this large family of  quasi-zero modes is responsible for the 
divergence of the variance of $A_{{\bm x},\mu}$ for $\mu < d$.

\subsection{Some other gauge fixings}

It is interesting to note that the results for the Lorenz gauge apply only to
the discretization (\ref{eq:Lorenzgauge}). If instead the discretization
(\ref{eq:Lorenzgauge2}) is used, different results are obtained.  Indeed, in
the latter case, in the infinite-volume limit, the gauge-fixed Hamiltonian is
invariant under a large family of gauge transformations.  For instance, one can
consider transformations like those reported in Eq.~(\ref{L2-example}).  To
determine the full  set of transformations that leave $F_{L',\bm x}(A)$
invariant in infinite volume, we work in Fourier space and consider a function
$\phi_{\bm x}$ of the form
\begin{equation}
   \phi_{{\bm x}} = a e^{i{\bm p}\cdot {\bm x}} + 
   \bar{a} e^{-i{\bm p}\cdot {\bm x}}, 
\end{equation}
where $a$ is an arbitrary complex constant. These transformations leave
$F_{L',\bm x}(A)$ invariant, if at least one of these two conditions is
satisfied:
\begin{equation}
\begin{aligned}
    & \sum_\mu \cos p_\mu (1 - \cos p_\mu) = 0, \\
    & \sum_\mu \sin p_\mu (1 - \cos p_\mu) = 0. 
   \label{LE-eq}
\end{aligned}
\end{equation}
If only the first (the second) equation is satisfied, then $a$ is necessarily 
real (purely imaginary).
The transformation (\ref{L2-example}) corresponds to taking 
${\bm p} = (\pi/2,-\pi/2,\pi/2,-\pi/2)$ and a real constant $a$. 
We have studied numerically the equations~(\ref{LE-eq}) in three dimensions, 
finding that both 
equations are satisfied on a two-dimensional surface in 
momentum space.
The presence of this family of 
gauge transformations that leave the Hamiltonian invariant, implies that
the correlation function $\widetilde{G}_{\mu\nu}(\bm p)$ is singular 
in ${\bm p}$ space. In turn, this implies (we have performed a numerical check) 
the divergence of 
the variance of $A_{\bm x}$, as it also occurs in the axial gauge.

Finally, we would like to make some comments on the Coulomb gauge that we 
can define as 
\begin{equation}
F_{L,{\bm x}}(A) = 
\sum_{\mu=1}^{d-1} (A_{{\bm x},\mu} - A_{{\bm x}-\hat{\mu},\mu}).
\label{eq:Landaugauge}
\end{equation}
In the hard case $\zeta = 0$, the correlation function is given by 
\begin{eqnarray}
\widetilde{G}_{\mu\nu} ({\bm p}) &=& 
{1\over \kappa} {\delta_{\mu\nu}\over \hat{p}^2} - 
    {1\over \kappa} {p_\mu p_\nu \over \hat{p}^2 \hat{p}_T^2} \qquad 
   \mu,\nu < d \nonumber \\
\widetilde{G}_{d\mu} &=&    0 \hphantom{xxxxxxxxxxxxxxxx}  \mu < d \nonumber \\
\widetilde{G}_{dd} &=& {1\over \kappa} {1\over \hat{p}_T^2}
\end{eqnarray}
where $\hat{p}_T^2 = \sum_{\mu=1}^{d-1} \hat{p}_\mu^2$. The susceptibilities
diverge as $L^2$ while $\langle B_{\bm x}\rangle$ is given by
\begin{equation}
\langle B_{\bm x} \rangle = {1\over \kappa} (2 I_{d,1} + I_{d-1,1}).
\end{equation}
In four dimensions, both sums are finite, therefore 
$\langle B_{\bm x} \rangle$ is well defined. In three dimensions, however, 
the result depends on the two-dimensional sum 
$I_{2,1}$, which diverges logarithmically. 
Therefore, for $d=3$, the photon mass operator is not well-defined in the 
Coulomb gauge.

\section{Ward identities in different gauges}\label{sec:B}

A crucial ingredient in the derivations presented in 
Sec.~\ref{sec:3}  are the  Ward identities satisfied by the 
correlation functions. 
We derive them here for the generic gauge-fixing function introduced 
in Sec.~\ref{sec:3}, see Eq.~(\ref{GF-M}). The corresponding function 
$S_{\rm GF}(A)$, defined in Eq.~(\ref{def-Gfunction}) is
given by
\begin{equation}
\begin{aligned}
S_M(A) &= \frac{1}{2\zeta V} \sum_{{\bm p}\,\alpha\beta} 
   M_{\alpha}({\bm p}) M_{\beta}(-{\bm p}) e^{-i(p_\alpha-p_\beta)/2}\times  \\ 
& \qquad \qquad \times \widetilde{A}_\alpha({\bm p}) \widetilde{A}_\beta(-{\bm p}).  
\end{aligned}
\end{equation}
Under an infinitesimal gauge transformation, we have 
\begin{equation}
\begin{aligned}
& \delta S_M = {1\over V} \sum_{\bm p} \delta_M({\bm p}) \widetilde{\phi} ({\bm p}), \\
& \delta_M({\bm p}) = 
       {1\over \zeta} \sum_{\alpha\beta} 
       M_{\alpha}({\bm p}) M_{\beta}(-{\bm p}) e^{-i(p_\alpha-p_\beta)/2}\times  \\
& \qquad \qquad
       \times (i \hat{p}_\alpha) \widetilde{A}_\beta(-{\bm p}).
\end{aligned}
\end{equation}
If we now consider $\langle A_{{\bm x},\gamma}\rangle$ and require its invariance under 
changes of variable represented by infinitesimal gauge transformations, we obtain
\begin{equation}
 \langle \Delta_\gamma \phi_{\bm x} + A_{{\bm x},\gamma} \delta S_M \rangle = 0.
\end{equation}
In Fourier space, this implies the relation
\begin{equation}
{1\over \zeta} \sum_{\alpha\beta}
       M_{\alpha}({\bm p}) M_{\beta}(-{\bm p}) e^{-i(p_\alpha-p_\beta)/2}
       \hat{p}_\alpha \widetilde G_{\gamma\beta}({\bm p}) = \hat{p}_\gamma.
\label{Ward}
\end{equation}
In the axial gauge we have $M_{\alpha}({\bm p}) = \delta_{\alpha d}$, while 
in the Lorenz gauge  we have $M_{\alpha}({\bm p}) = - e^{i p_\alpha/2} i \hat{p}_\alpha$. 
Substituting these relations in Eq.~(\ref{Ward}), we obtain 
Eqs.~(\ref{Ward-axial}) and (\ref{Ward-Lorenz}).


\begin{thebibliography}{99}

\bibitem{MM_book} 
  I. Montvay and G. M\"unster, 
  \textit{Quantum Fields on a Lattice}, 
  (Cambridge University Press, 1994).

\bibitem{DGDT_book}
  T. DeGrand and C. DeTar,
  \textit{Lattice methods for Quantum Chromodynamics}
  (World Scientific, 2006)

\bibitem{ZJ_book} 
  J. Zinn-Justin, 
  \textit{Quantum Field Theory and Critical Phenomena} 
  (Clarendon Press, 2002)

\bibitem{Pelissetto:2000ek}
  A. Pelissetto and E. Vicari,
  Critical phenomena and renormalization group theory,
  Phys. Rept. \textbf{368}, 549 (2002)

\bibitem{Sachdev_book}
  S. Sachdev,
  \textit{Quantum Phase Transitions} 
  (Cambridge University Press, 2011).

\bibitem{Seiler_book}
  E. Seiler, 
  \textit{Gauge theories as a problem of constructive quantum field theory and statistical mechanics}
  Lect. Notes Phys. 159 
  (Springer-Verlag, 1982).

\bibitem{GJ_book}
  J. Glimm and A.~Jaffe,
  \textit{Quantum Physics}
  (Springer-Verlag, 1987)

\bibitem{Gross:1973id}
  D. J. Gross and F. Wilczek,
  Ultraviolet Behavior of Nonabelian Gauge Theories,
  Phys. Rev. Lett. \textbf{30}, 1343 (1973).

\bibitem{Politzer:1973fx}
  H. D. Politzer,
  Reliable Perturbative Results for Strong Interactions?,
  Phys. Rev. Lett. \textbf{30}, 1346 (1973).

\bibitem{Coleman:1973sx}
  S. R. Coleman and D. J. Gross,
  Price of asymptotic freedom,
  Phys. Rev. Lett. \textbf{31}, 851 (1973).

\bibitem{MZ-03}
  M. Moshe and J. Zinn-Justin, 
  Quantum field theory in the large $N$ limit: A review, 
  Phys. Rep. {\bf 385}, 69 (2003).

\bibitem{Wilson:1973jj}
  K. G. Wilson and J. B. Kogut,
  The Renormalization group and the epsilon expansion,
  Phys. Rept. \textbf{12}, 75 (1974).

\bibitem{Bhanot:1980pc}
  G.~Bhanot and M.~Creutz,
  The Phase Diagram of $Z(n)$ and $U$(1) Gauge Theories in Three-dimensions,
  Phys. Rev. D \textbf{21}, 2892 (1980)

\bibitem{Caselle:2014eka}
  M.~Caselle, M.~Panero, R.~Pellegrini, and D.~Vadacchino,
  A different kind of string,
  J. High Ener. Phys. \textbf{01}, 105 (2015).

\bibitem{Athenodorou:2018sab}
  A.~Athenodorou and M.~Teper,
  On the spectrum and string tension of U(1) lattice gauge theory in 2 + 1 dimensions,
  J. High Ener. Phys. \textbf{01}, 063 (2019).

\bibitem{Teper:1998te}
  M. J. Teper,
  SU(N) gauge theories in (2+1)-dimensions,
  Phys. Rev. D \textbf{59}, 014512 (1999).

\bibitem{Athenodorou:2016ebg}
  A.~Athenodorou and M.~Teper,
  SU(N) gauge theories in 2+1 dimensions: glueball spectra and k-string tensions,
  J. High. Ener. Phys.  \textbf{02}, 015 (2017).

\bibitem{Bonati:2020orj}
  C.~Bonati and S.~Morlacchi,
  Flux tubes and string breaking in three dimensional SU(2) Yang-Mills theory,
  Phys. Rev. D \textbf{101}, 094506 (2020). 


\bibitem{SBSVF-04} 
  T. Senthil, L. Balents, S. Sachdev, A. Vishwanath and M. P. A.~Fisher, 
  Quantum Criticality beyond the Landau-Ginzburg-Wilson Paradigm, 
  Phys. Rev. B {\bf 70}, 144407 (2004).

\bibitem{Fradkin_book}
  E. Fradkin,
  \textit{Field Theories of Condensed Matter Physics}
  (Cambridge University Press, 2013).
 
\bibitem{Sachdev:2018ddg}
  S. Sachdev,
  Topological order, emergent gauge fields, and Fermi surface reconstruction,
  Rept. Prog. Phys. \textbf{82}, 014001 (2019).

\bibitem{Moessner_book}
  R.~Moessner, J.~E.~Moore,
  \textit{Topological Phases of Matter}
   (Cambridge University Press, 2021).

\bibitem{HLM-74} 
  B. I. Halperin, T. C. Lubensky, and S. K. Ma,
  First-Order Phase Transitions in Superconductors and Smectic-A Liquid Crystals, 
  Phys. Rev. Lett. {\bf 32}, 292 (1974).

\bibitem{FH-96} 
  R. Folk and Y. Holovatch, 
  On the critical fluctuations in superconductors, 
  J. Phys. A {\bf 29}, 3409 (1996).

\bibitem{IZMHS-19} B. Ihrig, N. Zerf, P. Marquard, I. F. Herbut, and 
M. M. Scherer, 
Abelian Higgs model at four loops, 
fixed-point collision and deconfined criticality, 
Phys. Rev. B {\bf 100}, 134507 (2019).

\bibitem{Das:2018qmx}
  A. Das,
Phase transition in 
$SU(N)\times U(1)$ gauge theory with many fundamental bosons,
Phys. Rev. B \textbf{97}, 214429 (2018).

\bibitem{Sachdev:2018nbk}
  S. Sachdev, H. D. Scammell, M. S. Scheurer and G. Tarnopolsky,
  Gauge theory for the cuprates near optimal doping,
  Phys. Rev. B \textbf{99}, 054516 (2019).

\bibitem{Bonati:2021tvg}
  C. Bonati, A. Franchi, A. Pelissetto and E. Vicari,
  Three-dimensional lattice SU($N_c$) gauge theories with multiflavor scalar fields in the adjoint representation,
  Phys. Rev. B \textbf{104}, 115166 (2021).

\bibitem{Bonati:2021rzx}
  C. Bonati, A. Franchi, A. Pelissetto and E. Vicari,
  Phase diagram and Higgs phases of three-dimensional lattice 
  SU($N_c$) gauge theories with multiparameter scalar potentials,
  Phys. Rev. E \textbf{104},  064111 (2021).

\bibitem{fermioni1}
J. Braun, H. Gies, L. Janssen, and D. Roscher,
  Phase structure of many-flavor QED$_3$, 
Phys. Rev. D {\bf 90}, 036002 (2014).

\bibitem{fermioni2}
I. F. Herbut, 
  Chiral symmetry breaking in three-dimensional quantum electrodynamics as fixed point annihilation, 
Phys. Rev. D {\bf 94}, 025036 (2016).

\bibitem{fermioni3}
  B. Ihrig, L. Janssen, L. N. Mihaila, and M. M. Scherer, 
  Deconfined criticality from the QED$_3$-Gross-Neveu model at three loops, 
  Phys. Rev. B {\bf 98}, 115163 (2018).

\bibitem{fermioni4}
  J. A. Gracey,
  Fermion bilinear operator critical exponents at $O(1/N^2)$ in the QED-Gross-Neveu universality class, 
  Phys. Rev. D {\bf 98}, 085012 (2018).

\bibitem{Bonati:2020jlm}
  C. Bonati, A. Pelissetto and E. Vicari,
  Lattice Abelian-Higgs model with noncompact gauge fields, 
  Phys. Rev. B \textbf{103}, 085104 (2021).

\bibitem{Bonati:2022ifi}
  C. Bonati and N. Francini,
  Noncompact lattice Higgs model with Abelian discrete gauge groups: Phase diagram and gauge symmetry enlargement,
  Phys. Rev. B \textbf{107}, 035106 (2023).

\bibitem{Bonati:2020ssr}
  C. Bonati, A. Pelissetto and E. Vicari,
  Higher-charge three-dimensional compact lattice Abelian-Higgs models,
  Phys. Rev. E \textbf{102}, 062151 (2020).

\bibitem{Bonati:2022oez}
  C. Bonati, A. Pelissetto and E. Vicari,
  Critical behaviors of lattice U(1) gauge models and three-dimensional Abelian-Higgs gauge field theory,
  Phys. Rev. B \textbf{105}, 085112 (2022)

\bibitem{Pelissetto:2019zvh}
  A. Pelissetto and E. Vicari,
  Three-dimensional ferromagnetic CP(N-1) models,
  Phys. Rev. E \textbf{100}, 022122 (2019).

\bibitem{Pelissetto:2019iic}
  A. Pelissetto and E. Vicari,
  Large-$N$ behavior of three-dimensional lattice CP$^{N-1}$ models,
  J. Stat. Mech. \textbf{2003}, 033209 (2020).
 
\bibitem{Pelissetto:2019thf}
  A. Pelissetto and E. Vicari,
  Multicomponent compact Abelian-Higgs lattice models,
  Phys. Rev. E \textbf{100}, 042134 (2019).

\bibitem{Bracci-Testasecca:2022mxc}
  G. Bracci-Testasecca and A. Pelissetto,
  Multicomponent gauge-Higgs models with discrete Abelian gauge groups,
 J. Stat. Mech.: Th. Expt. 043101 (2023).

\bibitem{MV-08} O.~I.~Motrunich and A.~Vishwanath, 
  Comparative study of Higgs transition in one-component and two-component lattice superconductor models,
  [arXiv:0805.1494 [cond-mat.stat-mech]] (unpublished).

\bibitem{KMPST-08} A.~B.~Kuklov, M.~Matsumoto, N.~V.~Prokof'ev, B.~V.~Svistunov, and M.~Troyer, 
  Deconfined Criticality: Generic First-Order Transition in the SU(2) Symmetry Case,
  Phys. Rev. Lett. {\bf 101}, 050405 (2008)
  [arXiv:0805.4334 [cond-mat.stat-mech]].

\bibitem{MV-04} 
  O. I. Motrunich and A. Vishwanath, 
  Emergent photons and transitions in the O(3) $\sigma$-model with hedgehog suppression, 
  Phys. Rev. B {\bf 10}, 075104 (2004).

\bibitem{MS-90} 
  G. Murthy and S. Sachdev, 
  Actions of hedgehogs instantons in the disordered phase of 
2+1 dimensional CP$^{N-1}$ model, 
  Nucl. Phys. B {\bf 344}, 557 (1990).

\bibitem{SP-15} 
  G. J. Sreejith and S. Powell, 
  Scaling dimensions of higher-charge monopoles at deconfined critical points, 
  Phys. Rev. B {\bf 92}, 184413 (2015).

\bibitem{Pelissetto:2020yas}
  A. Pelissetto and E. Vicari,
  Three-dimensional monopole-free $CP^{N-1}$ models,
  Phys. Rev. E \textbf{101}, 062136 (2020).

\bibitem{Bonati:2022srq}
  C. Bonati, A. Pelissetto and E. Vicari,
  Three-dimensional monopole-free CP$^{N-1}$ models: behavior in the presence of a quartic potential,
  J. Stat. Mech. \textbf{2206}, 063206 (2022).

\bibitem{BPV-in-prep} 
  C. Bonati, A. Pelissetto, and E. Vicari,
  in preparation.

\bibitem{Elitzur:1975im}
  S. Elitzur,
  Impossibility of Spontaneously Breaking Local Symmetries,
  Phys. Rev. D \textbf{12}, 3978 (1975).

\bibitem{DeAngelis:1977su}
  G.~F.~De Angelis, D.~de Falco and F.~Guerra,
  A Note on the Abelian Higgs-Kibble Model on a Lattice: Absence of Spontaneous Magnetization,
  Phys. Rev. D \textbf{17}, 1624 (1978). 

\bibitem{IZ_book}
  C.~Itzykson and J.~M.~Drouffe,
  \textit{Statistical field theory. vol. 1: from Brownian motion to renormalization and lattice gauge theory},
  (Cambridge University Press, 1989).

\bibitem{Kronfeld:1990qu}
  A. S. Kronfeld and U. J. Wiese,
  SU(N) gauge theories with C periodic boundary conditions. 1. Topological structure,
  Nucl. Phys. B \textbf{357}, 521 (1991).

\bibitem{Lucini:2015hfa}
  B. Lucini, A. Patella, A. Ramos and N. Tantalo,
  Charged hadrons in local finite-volume QED+QCD with C$^{*}$ boundary conditions,
  JHEP \textbf{02}, 076 (2016).

\bibitem{R_book}
  H. J. Rothe
  \textit{Lattice gauge theories. An introduction}
  (World Scientific, 2005).

\bibitem{Bonati:2021vvs}
  C. Bonati, A. Pelissetto and E. Vicari,
  Breaking of Gauge Symmetry in Lattice Gauge Theories,
  Phys. Rev. Lett. \textbf{127}, 091601 (2021)

\bibitem{Leibbrandt:1987qv}
  G.~Leibbrandt,
  Introduction to Noncovariant Gauges,
  Rev. Mod. Phys. \textbf{59}, 1067 (1987)

\bibitem{HT96}
  I. F. Herbut and Z. Te\v{s}anovi\'c
Critical Fluctuations in 
Superconductors and the Magnetic Field Penetration Depth,
  Phys. Rev. Lett. \textbf{76}, 4588 (1996).	

\bibitem{GZ-77} M. L. Glasser and I. J. Zucker,
  Extended Watson integrals for the cubic lattices,
  Proc. Natl. Acad. Sci. USA {\bf 74}, 1800 (1977).

\bibitem{JZ-01} G. S. Joyce and I. J. Zucker,
  Evaluation of the Watson integral and associated logarithmic integral
  for the $d$-dimensional hypercubic lattice,
  J. Phys. A: Math. Gen. {\bf 34}, 7349 (2001).

\bibitem{CP-98}
  S. Caracciolo and A. Pelissetto,
  Corrections to Finite-Size Scaling in the Lattice $N$-Vector Model
  for $N=\infty$,
  Phys. Rev. D {\bf 58},105007 (1998).

\end{thebibliography}
\end{document}